\documentclass[twocolumn]{aastex63}

\usepackage{graphicx}
\graphicspath{ {images/} }
\usepackage{amsmath,amssymb}
\usepackage{color}
\usepackage{units}
\usepackage{hyperref}
\newcommand\note[1]{\textcolor{red}{(note: #1)}}
\newcommand\pdot{\dot{P}}
\newcommand\cosmic{\texttt{COSMIC}}

\providecommand{\keywords}[1]
{
  \small	
  \textbf{\textit{Keywords---}} #1
}

\begin{document}

\title{GPU-Accelerated Periodic Source Identification in Large-Scale Surveys: Measuring $P$ and $\pdot$}

\author{Michael L. Katz}
\affiliation{Department of Physics and Astronomy, Northwestern University, Evanston, IL 60208, USA}
\affiliation{Center for Interdisciplinary Exploration and Research in Astrophysics (CIERA), Evanston, IL 60208, USA}

\author{Olivia R. Cooper}
\affiliation{Department of Astronomy, Smith College, Northampton, MA 01063, USA}

\author{Michael W. Coughlin}
\affiliation{School of Physics and Astronomy, University of Minnesota, Minneapolis, Minnesota 55455, USA}
\affiliation{Division of Physics, Math, and Astronomy, California Institute of Technology, Pasadena, CA 91125, USA}

\author{Kevin B. Burdge}
\affiliation{Division of Physics, Math, and Astronomy, California Institute of Technology, Pasadena, CA 91125, USA}

\author{Katelyn Breivik}
\affiliation{Canadian Institute for Theoretical Astrophysics, University of Toronto, 60 St. George Street, Toronto, Ontario, M5S 1A7, Canada}

\author{Shane L. Larson}
\affiliation{Department of Physics and Astronomy, Northwestern University, Evanston, IL 60208, USA}
\affiliation{Center for Interdisciplinary Exploration and Research in Astrophysics (CIERA), Evanston, IL 60208, USA}

\date{\today}

\begin{abstract}

Many inspiraling and merging stellar remnants emit both gravitational and electromagnetic radiation as they orbit or collide. These gravitational wave events together with their associated electromagnetic counterparts provide insight about the nature of the merger, allowing us to further constrain properties of the binary. With the future launch of the Laser Interferometer Space Antenna (LISA), follow up observations and models are needed of ultracompact binary (UCB) systems. Current and upcoming long baseline time domain surveys will observe many of these UCBs. We present a new fast periodic object search tool capable of searching for generic periodic signals based on the Conditional Entropy algorithm. This new implementation allows for a grid search over both the period ($P$) and the time derivative of the period ($\pdot$). To demonstrate the usage of this tool, we use a small, hand-picked subset of a UCB population generated from the population synthesis code \cosmic, as well as a custom catalog for varying periods at fixed intrinsic parameters. We simulate light curves as likely to be observed by future time domain surveys by using an existing eclipsing binary light curve model accounting for the change in orbital period due to gravitational radiation. We find that a search with $\pdot$ values is necessary for detecting binaries at orbital periods less than $\sim$10 min. We also show it is useful in finding and characterizing binaries with longer periods, but at a higher computational cost. Our code is called \texttt{gce} (GPU-Accelerated Conditional Entropy). It is available on \href{https://github.com/mikekatz04/gce/tree/c3495ed7c1316cc26bb7362417d95d7832671c6f}{Github}.$^1$  
\end{abstract}

\footnotetext{\href{https://github.com/mikekatz04/gce/tree/c3495ed7c1316cc26bb7362417d95d7832671c6f}{https://github.com/mikekatz04/gce}}


\section{Introduction}
Our galaxy is rich with a menagerie of binary objects, many of which evolve into dense stellar remnants rapidly orbiting each other in ultracompact binary systems (UCBs). These binaries can be detached or interacting, and are characterized by periods of one hour or shorter \citep{2013nelemansUCB}. UCBs provide insight into many poorly understood stellar processes including common-envelope evolution \citep[e.g.][]{Woods2012DWD_common_env, McNeill2020_GW_tides_DWD}, magnetic braking, and massive star evolution \citep{2013nelemansUCB}. 

UCBs also emit gravitational waves (GW) in the mHz frequency regime. The future space-based GW detector, the Laser Interferometer Space Antenna (LISA), is highly sensitive to UCBs and other objects \citep[e.g.][]{2012nissanke, 2013nelemansUCB, LISA2017, 2017korol, Lamberts2019LISA_DWD, Breivik2019, Korol2020Populations_DWD, Lau2020BNS_LISA}. LISA is a future mission by the European Space Agency (ESA) with NASA as a mission partner. LISA is slated for launch in the early 2030s and is expected to detect $>10^4$ slowly inspiraling UCBs \citep{LISA2017}. To understand the potential of future GW and multi-messenger observations of UCBs, we can use optical data to identify and analyze UCBs in the electromagnetic (EM) regime. 

Double white dwarfs (DWDs), a type of UCB, are very common in our Galaxy as the majority of stars evolve into white dwarfs \citep{2017korol}. DWDs are found in two main configurations: detached or semi-detached (AM CVn). Detached systems are typically identified by eclipses in their light curves \citep{2017korol}. Semi-detached AM CVn, on the other hand, are found using spectral features; specifically, AM CVns exhibit accretion features and He lines in their spectra \citep{2012nissanke}. The merger of these DWD systems may be progenitors to rare, massive white dwarfs, solo neutron stars, subdwarf-O stars, or R Corona Borealis stars \citep{2012nissanke}. DWDs in a sub-hour binary can also help in the study of tides, white dwarf internal characteristics, and white dwarf viscosity \citep[e.g.][]{Kremer2015, 2017korol, Breivik2018}. 


LISA Verification Binaries are DWD systems that have been found with EM observations to have orbital frequencies $\sim$mHz, indicating these systems will emit GWs observable by LISA. Approximately 10 systems detected to date are predicted to reach a detectable signal-to-noise ratio (SNR) threshold over the duration of the LISA mission \citep{LISAbinaries2018}. The current sample in \citet{LISAbinaries2018} exhibit sources mostly in the Northern hemisphere at high Galactic latitudes, suggesting the sample is incomplete. Optical surveys such as those utilizing the Zwicky Transient Facility \citep[ZTF;][]{2019ZTFsciobj} and the Vera C. Rubin Observatory's \citep[VRO;][]{VRO2020} Legacy Survey of Space and Time \citep[LSST;][]{2019LSST} should discover new sources building a more complete sample of Verification Binaries. However, many of the degenerate LISA UCBs are inherently faint \citep[up to 70th mag;][]{2017korol} and therefore have not been and will not be identified with EM observations. Still, \citet{2017korol} estimate $\sim2 \times 10^3$ eclipsing, detached DWDs will be detectable by VRO.

Not every light curve identified as being significantly periodic, however, will correspond to a DWD binary source. The sample will be contaminated with, for example, pulsating white dwarfs, Delta Scuti variables, SX Phoenicis stars, and cool spots on the stellar surfaces coming in and out of view as the star rotates \citep{2017korol}. Analysis of the shape of the light curve \citep{burdge7min} as well as the object's position near the white dwarf track on an HR diagram will allow us to identify the DWD systems we are interested in \citep{coughlin40min}.

To find periodic sources, we use a variety of search methods within large-scale surveys. For details of this process and a comparison of methods, please see \citet{Graham2013period_finding}. For example, as detailed in \citet{burdge7min}, for searches for UCBs with ZTF, period searches are performed on light curves, and then sources of interest are followed up with high cadence photometric instruments such as Chimera \citep{HaHa2016} or the Kitt Peak Electron Multiplying CCD (KPED) \citep{KPED2019}. 

To support efforts of this type, in this paper, we introduce our new search tool: \texttt{gce} (GPU-Accelerated Conditional Entropy). The purpose of this tool is to search large optical catalogs for periodic objects of interest using the Conditional Entropy (CE) technique \citep{Graham2013}. The CE technique is general enough to search for all forms of periodic phenomena. However, we focus on DWD binaries as an example of how our new tool is necessary for accomplishing broad astrophysical studies with these sources. With this in mind, the speed metrics detailed below will apply to any source, not just DWD systems. 

When identifying periodic objects with the CE, period ($P$) values are searched over a grid; a CE value is calculated at each grid point. The lowest Conditional Entropy value indicates the most likely period (we will explain this process in more detail in \mbox{Section \ref{sec:methods-ce}}). The immense speed and efficient memory usage of our algorithm gives us an important benefit. For the first time, we can extend beyond a one-dimensional grid over period values to a two-dimensional grid of the period and the time derivative of the period ($\pdot$). This ability is critical for identifying periodic sources in short-period orbital configurations exhibiting strong orbital decay. Sources that exhibit a non-negligible orbital decay rate, like short-period DWDs, will become undetectable with period search algorithms if the $\pdot$ of the system is not accounted for. We refer to this as ``dephasing'' based on the phase-folding method described in \mbox{Section \ref{sec:methods-ce}}. In the Pulsar community, $\pdot$s are used in a similar way as a flexible diagnostic tool \citep[e.g.][]{Tauris:2001cy, Johnston:2017wgm}.

Separate from purely being able to observe these systems, $\pdot$s are astrophysically interesting. According to general relativity, the orbital evolution of DWDs with characteristically short periods is primarily driven by GW radiation \citep{2013nelemansUCB}. This orbital decay manifests in a decreasing orbital period over time, given by \citep{Peters1963},
\begin{equation}\label{eq:gwrad}
    \pdot_\text{gw} = -(2)\frac{96\pi}{5c^5}\left(\frac{2G\pi M_{c}}{P_\text{orb}}\right)^{5/3},
\end{equation}
 where $M_{c}$ is the chirp mass. To be clear, $\pdot_\text{gw}$ represents the \textit{orbital} decay rate due to gravitational radiation, not the time derivative of the period of the gravitational wave itself. However, certain astrophysical scenarios, like DWD direct impact accretion, will produce $\pdot$ values that will vary compared to $\pdot_\text{gw}$ \citep{Kremer2015}. With $\pdot_\text{obs}$ measurements (``obs'' indicates observed) from EM observations with our algorithm, or LISA observations, the underlying astrophysical scenario can be illuminated because $\pdot_\text{obs} = \pdot_\text{astro} + \pdot_\text{gw}$ \citep{Breivik2018}. Interestingly, in certain scenarios, $\pdot$ can even be greater than zero, indicating the orbital separation is increasing \citep{Kremer2015, Breivik2018}.  
 
In the following, we will detail the DWD groups analyzed and the simulation of their light curves to perform this study. Additionally, we will discuss the implementation of our CE algorithm, including specific aspects of our code that result in the performance increase \mbox{(Section \ref{sec:methods})}. We will then detail the recovery of these simulated sources using \texttt{gce} \mbox{(Section \ref{sec:results})} and discuss implications of the use of such an algorithm in future EM observations \mbox{(Section \ref{sec:discuss})}.

\section{Methods}\label{sec:methods}
\subsection{Generating the Ultra-Compact Binary Population}\label{sec:genpop}

For this work, we analyze two groups of DWD binaries. The first is a group of binaries with periods ranging from 4 to 38 minutes in steps of 1 minute. We assume these binaries exhibit a $\pdot$ determined solely by GWs for simplicity. The purpose of this group is to understand the behavior of our algorithm over a range of periods and their associated GW-based $\pdot$ values. We specifically chose to fix other parameters to the mean values measured for the 7 minute binary DWD from \citet{burdge7min}, ZTF J1539+5027, in order to restrict the parameter space for this data set. The primary and secondary masses were set to 0.61$M_\odot$ and 0.21$M_\odot$ respectively. The radii for each of these masses was set to 0.0156$R_\odot$ and 0.0314$R_\odot$ respectively. The inclination was set to 84.2$^\circ$. Finally, we set the magnitude to 16. This magnitude value represents a brighter source than the DWD from \citet{burdge7min}. This value was chosen to ensure longer period binaries with those parameters still exhibited a measurable eclipse when applying photometrically appropriate errors to the magnitude value of each observation. The errors are large enough at $m\sim19$ to drown out the signal at longer periods. We will refer to this data set as the ``Burdge'' set.

To simulate a realistic subset of a galactic population of DWDs, we use \cosmic$\ $\citep{COSMICcode, Breivik2019}. This allowed us to test our algorithm on binaries with a variety of parameters rather than gridding parameters along multiple axes beyond just the period axis analyzed with our Burdge group. Within \cosmic$\ $we employed a ``DeltaBurst'' star formation setting and assumed an initial metallicity of 0.02. Other \cosmic$\ $parameters chosen for the population synthesis, including settings for binary evolution, can be found in our \href{https://github.com/mikekatz04/gce/tree/c3495ed7c1316cc26bb7362417d95d7832671c6f}{Github} repository. \cosmic$\ $ initially generates a statistically representative sample of DWDs based on the chosen settings. For this step, \cosmic$\ $produced 72,837 binaries. We then sample from this population so as to scale by mass to the mass estimates of the Thin Disk by \citet{McMillan2011GalPop}. This turned our initial sample into $\sim8\times10^8$ DWD binaries. These binaries are then spatially arranged appropriately in the Thin Disk \citep{McMillan2011GalPop} and evolved in terms of their temperature and luminosity to the present day using Equation 90 from \citet{Hurley2000StellarEvo} based on \citet{Mestel1952WDs}. With this present day luminosity, we determine the apparent magnitude of each binary based on its assigned spatial position relative to our Sun placed at a radius of 8 kpc \citep{Boehle2016SagADistance} from the galactic center. A cut is then made based on the dimmest magnitude of ZTF \cite[20.5;][]{2019ZTFsystemoverview}, as well as requiring periods under one hour as these systems are a primary focus for LISA, as well as sources that exhibit stronger and, therefore, more observable orbital decay. These cuts reduce our galactic population from $\sim10^8$ binaries to $\sim10^4$ binaries. We point the reader to the \cosmic$\ $\href{https://cosmic-popsynth.github.io/docs/latest/}{documentation} for more information on sampling a galactic population of DWD binaries.

\begin{figure*}
\begin{center}
\includegraphics[scale=0.5]{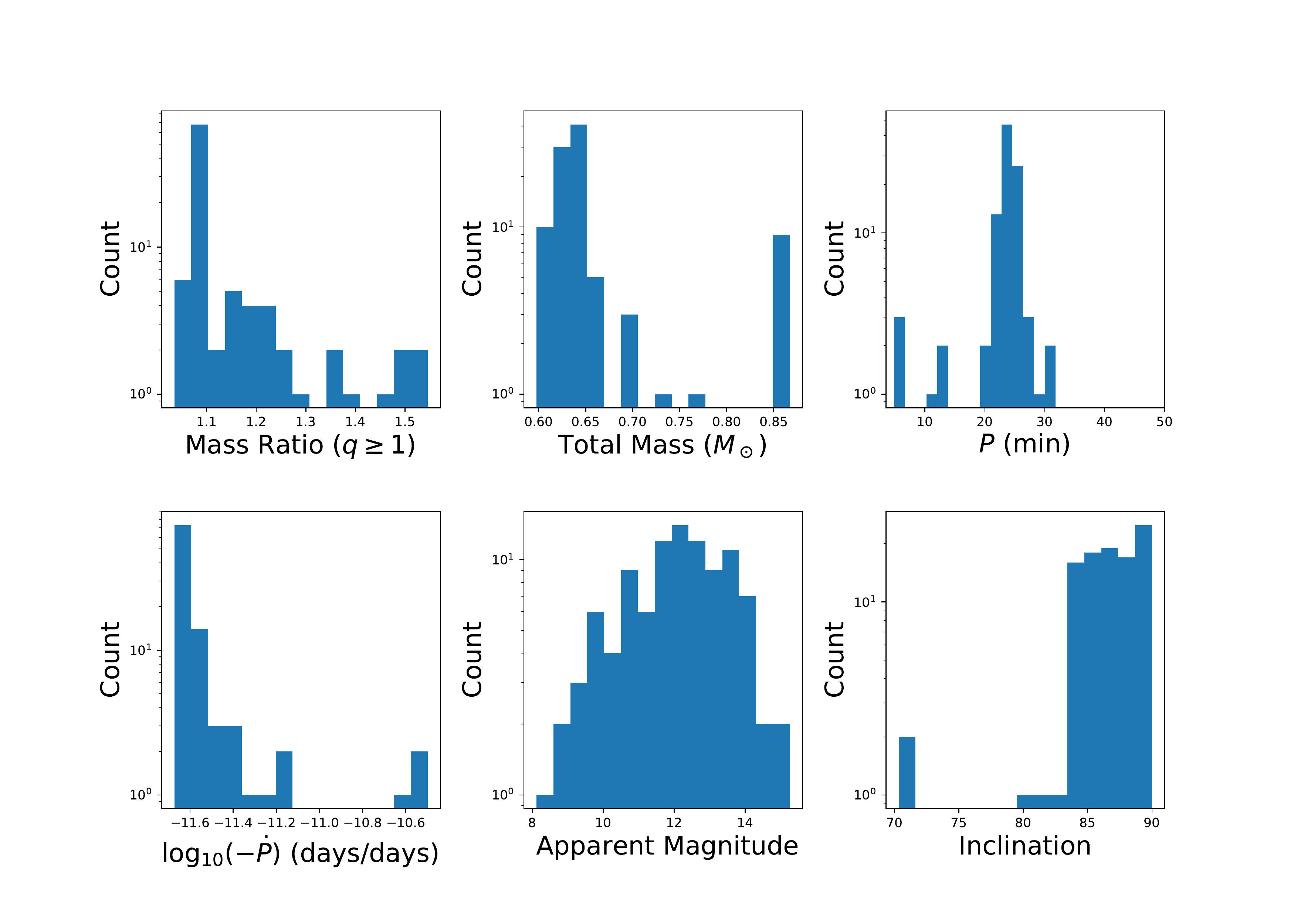}
\caption{Histograms for the main parameters of binaries within our \cosmic$\ $catalog are shown above. The distributions shown here represent the catalog we actually analyzed after performing magnitude-based and period limiting cuts. The mass ratio, total mass, and period are shown in the top row from left to right. The total mass and period are represented in solar masses and minutes, respectively. The bottom row displays the $\pdot$, magnitude, and inclination from left to right. The inclination is shown in degrees where 90$^\circ$ represents an ``edge-on'' configuration. Since the $\pdot$s we analyze are negative, and span over an order of magnitude, we present the $\pdot$ histogram as the log$_{10}$ of the negative of the $\pdot$ value.}\label{fig:population}
\end{center}
\end{figure*}

We then made cuts based on the light curve (see \mbox{Section \ref{sec:genlcs}}) to ensure we were running our algorithm on detectable sources. Specifically, we required a dimmest magnitude of 22 (after measurement errors were included), just beyond the detection limit for ZTF, and an eclipse depth (median magnitude minus the dimmest magnitude) of 0.2. Ultimately, we were left with $\sim500$ binaries. We wanted to analyze a sufficient number of binaries while maintaining a manageable number in order to focus on the algorithmic performance rather than analyzing the full detection prospects for a specific telescope with a full population of sources. This could be a topic for future work. For this reason, we chose to analyze the one hundred shortest period binaries from the $\sim500$ binaries that make it through all of our cuts.
Parameters for our final set of binaries can be seen in \mbox{Figure \ref{fig:population}}. Both this catalog generated with \cosmic, as well as our Burdge catalog, and their associated light curves are available in the \href{https://github.com/mikekatz04/gce/tree/c3495ed7c1316cc26bb7362417d95d7832671c6f}{Github} repository.

\subsection{Simulating Light Curves with Orbital Decay}\label{sec:genlcs}

To simulate light curves with orbital decay, we used the \texttt{ellc} package \citep{ellc} to create eclipsing binary light curves with a constantly changing period. The parameters we input into \texttt{ellc} include $\{m_1, m_2, r_1, r_2, m, P, \pdot, \iota, s/b\}$, where $m_1$ and $m_2$ are the masses of the two WDs; $r_1$ and $r_2$ are the respective radii of the two WDs; $m$ is the apparent magnitude of the system; $\iota$ is the inclination of the orbital plane; and $s/b$ is the surface brightness ratio of the secondary to the primary. We fix the surface brightness ratio for each binary to 0.5 for simplicity. This value was chosen to resemble an average binary similar to the DWD system from \citet{coughlin40min}. We point the reader to \citet{ellc} for specifics on the constructions used for eclipsing binary light curves. To represent the orbital decay, we generate a light curve for a new period at each observation time and then stitch these together ensuring the orbit achieves the proper phase at each time. To construct a single output light curve for the series of light curves with decreasing periods, we compute the modulus of the time for each observation and corresponding $P(t)$, and interpolate at the desired input times to achieve an eclipsing light curve exhibiting orbital decay. Over the baselines we are considering, which is $\sim10$ years for observations of DWD systems, we can focus on the initial period ($P_0$) and $\pdot$ (we do not have to consider $\ddot{P}$ due to its negligible influence on the timescales considered):
\begin{equation}\label{eq:P(t)}
    P(t) = P_0 + \pdot t + \mathcal{O}(t^2),
\end{equation}
where $P_{0}$ is in units of time and $\pdot$ is in units of time/time. General relativity predicts that a typical DWD system with $\mathcal{M}_c$ (chirp mass) on the order of $\sim0.5 M_{\odot}$ and a sub-hour period will have a negative $\pdot$ with magnitude on the order of \mbox{$\sim10^{-13}-10^{-11}$ ss$^{-1}$} \citep{Peters1963}.

To simulate realistic time sampling, we generate time arrays of length $n$. The time difference between each observation point is sampled from a Gaussian distribution with mean $\Delta t$ set appropriately according to the survey being simulated. For example, we examine VRO-like \citep{VRO2020} light curves employing a baseline of 10 years with approximately 7 days between each observation. Therefore, time sampling parameters were set to $n\sim500$ and $\Delta t = 7$.

The photometric data are taken to be Poisson noise dominated. We assume the observed magnitude is normally distributed about the true magnitude with an intrinsic standard deviation dependent on the filter and source magnitude. We use data direct from ZTF for the intrinsic standard deviation as a function of magnitude in $g$-band and $r$-band. These values are interpolated to calculate realistic error bars for a given light curve. Note that this ZTF-based error is about an order of magnitude higher than photometric errors predicted for VRO. For an object with $r$-mag = 21, \citet{2009LSSTsciencebook} report a photometric error of $\sigma = 0.01$ mag for VRO, while we use $\sigma \sim 0.2$ for ZTF-like photometric errors. Therefore, our light curves are a conservative estimate of the observed light curves by VRO. For an extensive discussion on ZTF photometric errors, please see \citet{MasciZTFErrors2019}.

\subsection{Conditional Entropy Implementation}\label{sec:methods-ce}

To prepare for the CE calculation, we normalize the magnitude values for each specific light curve. To normalize, we choose to subtract from each observation the median magnitude value, and then we divide by the standard deviation of the magnitude values. One benefit of this method is this allows us to combine observations from different filters into one input light curve for the CE calculation. Empirically, this method generally enhances the ability to find the correct system parameters. However, with that said, we do not examine different filters in this work for simplicity. 
    
\begin{figure*}
\begin{center}
\includegraphics[scale=0.5]{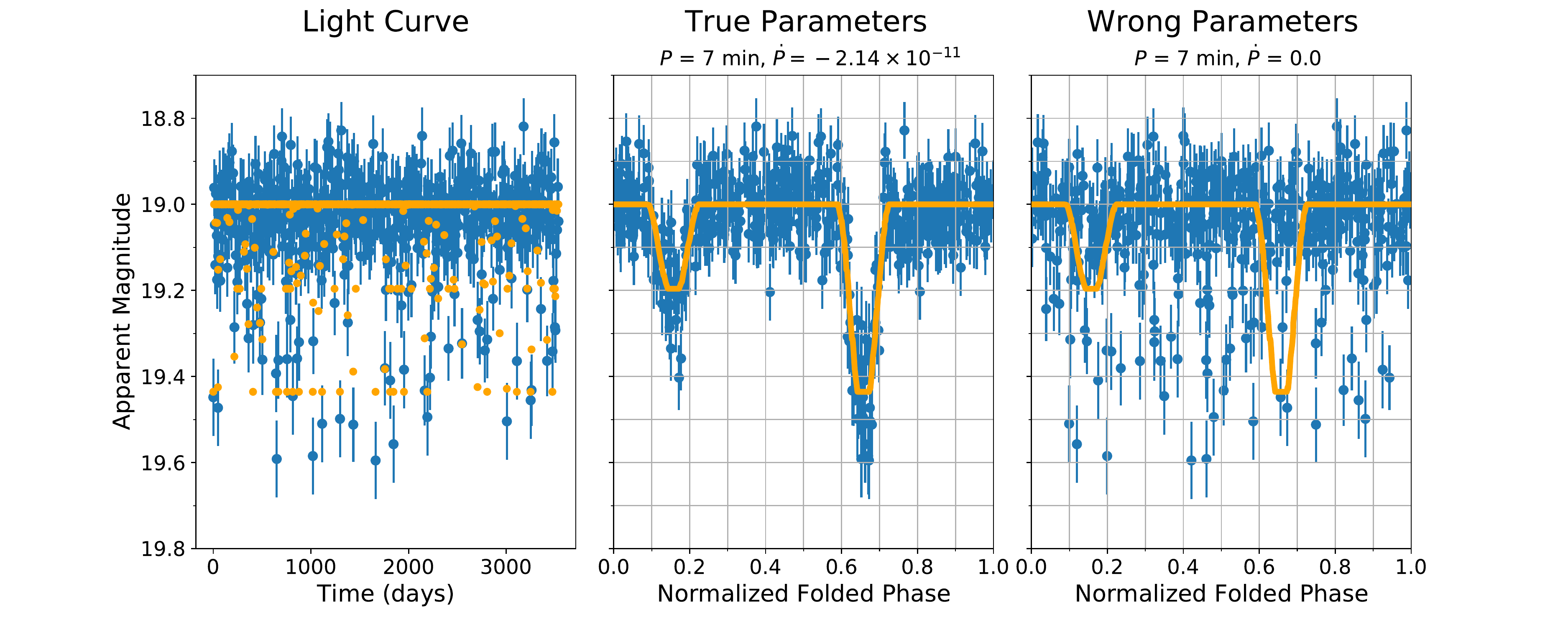}
\caption{Example light curves are shown above for an eclipsing binary similar to ZTF J1539+5027 \citep{burdge7min}. The true period and $\pdot$ of this binary is 7\,min and $-2.14\times10^{-11}$ss$^{-1}$, respectively. The light curves shown in blue contain errors. The base template for the light curves without any errors is shown in orange. The left plot shows the light curves as observed over our 10 year baseline at an average of once per week (see \mbox{Section \ref{sec:genlcs}}). The central plot presents the phase-folded light curve with the true period and $\pdot$. With the correct parameters, the eclipsing structure is clearly observed. The right plot shows an example of phase-folding with incorrect parameters. In this case, we show the phase-folded light curve using a period of 7 min and a $\pdot$ of zero. These parameters were chosen purposefully to illustrate the difference between searching for these parameters with and without a $\pdot$ axis. Grid lines are added to the middle and right plots to give the reader a sense of the binning technique used in the Conditional Entropy; however, this is only for example as it does not represent the true binning settings chosen for this work.}\label{fig:lc_ex}
\end{center}
\end{figure*}

With a normalized light curve in hand, we choose $P$ and $\pdot$. These values are taken from our two-dimensional $n_P$ x $n_{\pdot}$ grid set as the input to our algorithm. The time values are then phase-folded, given by,
\begin{equation}\label{eq:phase_fold}
    P\phi(t) =\left(t - \frac{1}{2} \frac{\pdot}{P}t^2 \right) \texttt{mod}\  P,
\end{equation}
where $\phi$ is the phase-folded value of the observation time with a range from 0 to P and \texttt{mod} is the modulus operator. We then divide both sides by $P$ to get the normalized phase-folded value ranging from 0 to 1. An example light curve at $m=19$ and the process of phase folding are shown in \mbox{Figure \ref{fig:lc_ex}}. The example shows the visual arrangement of light curve points when folded with both the correct (central plot) and incorrect (right plot) combination of $P$ and $\pdot$. With the correct parameters ($P=7$ days, $\pdot=-2.14\times10^{-11}$ days/days), the folded light curve presents the structure of two eclipses separated by half of the period. For the incorrect parameters, we chose to represent the correct period ($P=7$ days), but a $\pdot$ of zero to show the effect of not searching over $\pdot$ in addition to the period. It is clear in this case that the folded light curve does not contain any visual structure indicating we are near the proper values.

We implement the CE as described in \citet{Graham2013}. These normalized magnitude and phase values are then binned in a two-dimensional histogram with $n_\text{mag}$ magnitude bins and $n_\text{phase}$ phase bins. We have designed our implementation so that each histogram bin has a 50\% overlap with each adjacent bin. We choose $n_\text{mag}=10$ and $n_\text{phase}=50$ for the eclipsing systems analyzed in this paper. With the binned data we apply the CE, also referred to as $H_c$, given by \citep{Graham2013},
\begin{equation}\label{eq:ce}
    H_c = \sum_{i,j}p(m_i, \phi_j)\ln{\left( \frac{p(\phi_j)}{p(m_i, \phi_j)} \right)},
\end{equation}
 where $i,j$ are the index representing the $i$th magnitude bin and the $j$th phase bin. $p$ is the probability that an observed point is located in the bin with index $\{i,j\}$. 
 
Once we have calculated the CE for all $P$ and $\pdot$ configurations and have found the minimum value, we want to quantify the significance of this minimum value in comparison to other values tested. With this in mind, we choose to quantify our significance, $\rho$, given as,
\begin{equation}\label{eq:signif}
    \rho = \frac{\mu - H_{c,\text{min}}}{\sigma}, 
\end{equation}
where $\mu$ and $\sigma$ represent the mean and standard deviation of all CE measurements for a singular light curve. This effectively quantifies how different the minimum CE entropy value is from the pack of CE values that do not resemble the true parameters. Since the CE is relative to each light curve, we use the mean and standard deviation as an effective normalization. 

In the original formulation of the CE, the errors in the observations are not included. We considered ways of adding this to our calculation, but ultimately decided not to include this change. The observed points during the eclipsing phase of the system exhibit lower magnitudes due to the nature of an eclipsing system. These lower magnitudes lead to larger error bars. Generally, including errors involved weighting points based on the relative error for each observed point. With these eclipsing sources, we are most concerned about catching observations during the eclipsing phase of the orbit. Therefore, weighting our observations based on the error would down-weight observations from the eclipsing phase, directly counter to our goal.

\subsubsection{$\pdot$ Resolution}

The CE will be sensitive to $\pdot\neq0$ after observing a system for time $t$ if the change in the normalized phase when including $\pdot \neq0$ compared to $\pdot=0$, $\Delta\Phi$, becomes the same order as the bin width. Since we have normalized our phase-fold from 0 to 1, the bin width is equivalent to the inverse of the number of bins, $1/n_b$. Therefore, from \mbox{Equation \ref{eq:phase_fold}},
\begin{equation}
    \Delta\Phi = \frac{1}{2}\frac{|\dot{P}|}{P^2}t^2  \gtrsim \frac{1}{n_b}.
\end{equation}
The limiting detectable initial period, $P^*$, is then given by,
\begin{equation}
    P^*(t) \lesssim \left(\frac{1}{2}n_b |\pdot|\right)^{1/2}t.
\end{equation}
Assuming $\pdot=\pdot_\text{GW}$ from \mbox{Equation \ref{eq:gwrad}}, we find,
\begin{equation}
     P^*(t) \lesssim \left(n_b \frac{96\pi}{5c^5}\right)^{5/3}\left(2G\pi M_{c}\right)^{5/11} t^{6/11}.
\end{equation}
With the parameters from ZTF J1539+5027, 50 phase bins, and an observation time of 10 years, $P^*\sim30-40$ min, representing the higher end of our catalog period range. Please note that this calculation of $P^*$ does not consider the actual detectability or significance of a specific source. Furthermore, this calculation does not answer the question of whether a source is detectable with $\pdot=0$. We strive to answer this question in \mbox{Section \ref{sec:res_sig}}.

\subsubsection{GPU Implementation}\label{sec:gpu_impl}

The previous iteration of CE used in similar work was from the \texttt{cuvarbase} package \citep{cuvarbase}. Like \texttt{gce}, \texttt{cuvarbase} is implemented on Graphics Processing Units (GPU). The key development that allowed for the speed increase with \texttt{gce} is related to memory considerations and GPU architecture choices. The actual CE algorithm is effectively the same. 

The CE algorithm in \texttt{gce} is implemented directly in \texttt{CUDA} \citep{CUDA}. GPU kernels are programmed based on three-dimensional grids of blocks, where each block can launch a three-dimensional set of threads. The threads are the programming piece that actually run the provided code. Therefore, effectively utilizing blocks and threads are essential to maximize computing performance. For more information on GPU programming, please see the \texttt{CUDA} documentation. 

 \texttt{cuvarbase} allocated one block for each $P$ value from the one-dimensional grid (\texttt{cuvarbase} was unable to handle $\pdot$ in a tractable amount of time due to speed and memory considerations). It then used shared memory and sets of threads to efficiently phase-fold and form the two-dimensional histogram. With this two-dimensional histogram stored in shared memory, it would calculate the CE value. For more specific details of the \texttt{cuvarbase} implementation, we refer the reader to its documentation. The key limitation of  \texttt{cuvarbase} is that it uses its main parallelism to create a singular histogram. The change we made allows for each thread, rather than each block, to run its own $P$ (and $\pdot$) grid point. In order to accomplish this level of parallelism, the fundamental realization is when calculating the CE according to \mbox{Equation \ref{eq:ce}}, the only memory that needs to be stored is the phase bin counts corresponding to a singular magnitude value. This effectively means it is possible to deal, in terms of memory, with one-dimensional phase histograms, rather than an entire two-dimensional histogram. Therefore, the amount of necessary memory per $P$ (and $\pdot$) point is approximately $1/n_\text{mag}$ less than the \texttt{cuvarbase} implementation. 
 
\begin{figure}[h]
\begin{center}
\includegraphics[scale=0.3]{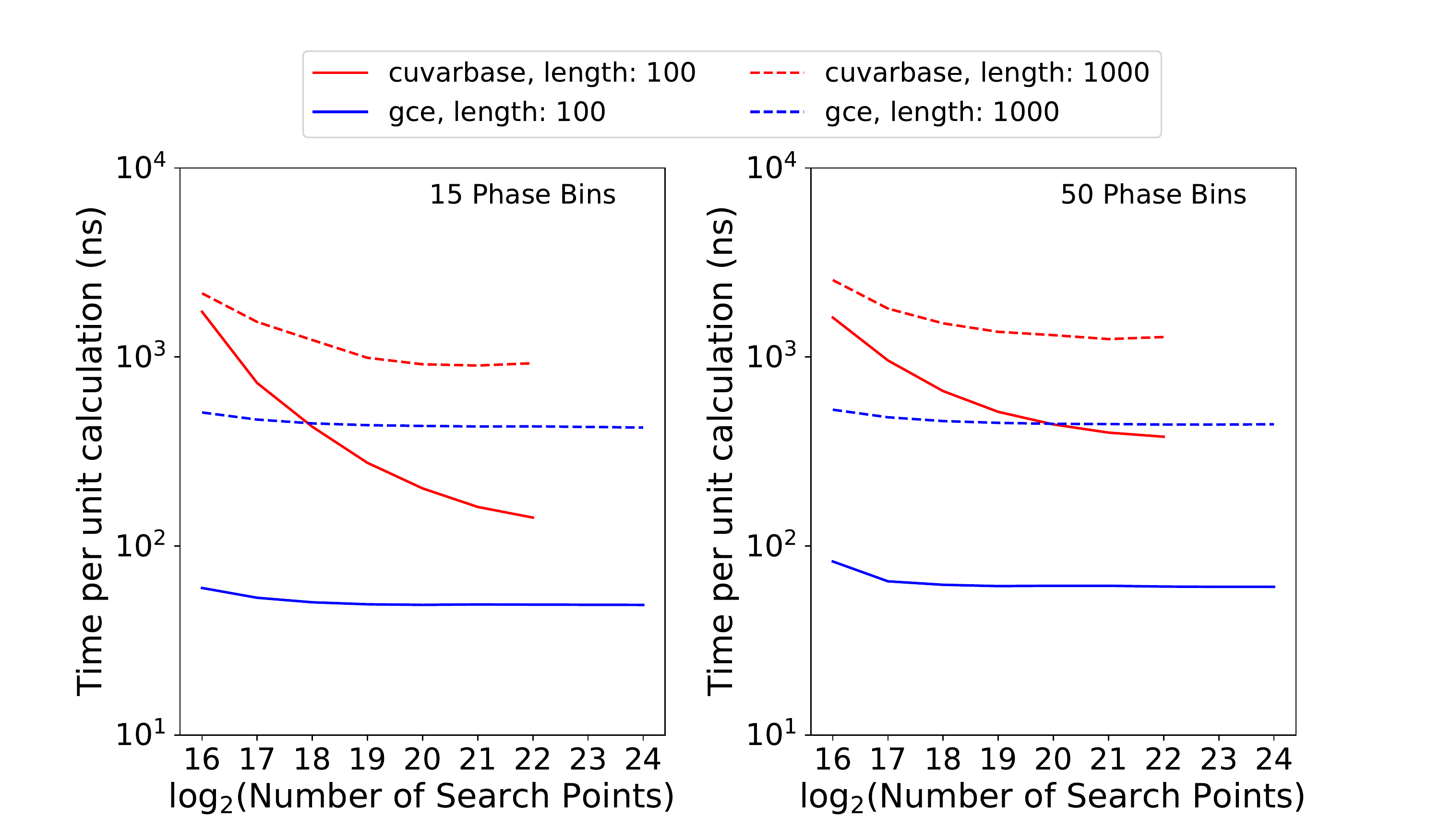}
\caption{The time scaling of our algorithm is shown above. The scaling is shown as the time per light curve per search point, meaning that we take the total timing of a run over all light curves and search points and normalize to the unit calculation. The term ``search points'' is used to represent the combination of all $P$ and $\pdot$ grid points over which we search. As long as the number of search points is sufficiently high, there is no difference in per unit timing when adding an additional period or $\pdot$ to the search grid. We represent four cases to give a sense of performance in a variety of situations. Phase bin counts of 15 and 50 are shown in blue and red, respectively. In our analysis, we use 50 phase bins that overlap by 50\% with surrounding bins. This is a typical number used for eclipsing binaries. When examining more sinusoidal phenomena, 15 bins can be used. In addition to the phase bin settings, the length of the light curves analyzed are shown: solid (dashed) lines represent light curves of 100 (1000) points. As the lines tend toward zero slope at higher numbers of points, the GPU experiences saturation indicating its cores are all busy throughout a majority of the calculation. At this point the timing gain slows down.}\label{fig:timing}
\end{center}
\end{figure}

 Our CE algorithm is wrapped to \texttt{Python} using \texttt{Cython} \citep{Cython} and a special \texttt{CUDA} wrapping code from \cite{CUDAwrapper}. \texttt{gce} itself is a \texttt{Python} code that prepares light curves and their magnitude information for input into the \texttt{Cython}-wrapped \texttt{CUDA} program. Therefore, the user interface is entirely \texttt{Python}-based. We provide two CE search functions based on the length of the light curves being analyzed. One is an optimized implementation for shorter light \mbox{curves ($\lessapprox$ 1400 points)}. In this work, we exclusively use this optimized version. We also provide a base implementation of the CE that can handle any length light curve albeit at the cost of efficiency for shorter light curves. Options for batching in groups of light curves and/or $\pdot$s are provided to help alleviate memory capacity issues. Timing for our algorithm with different amounts of phase bins and light curve points is shown in \mbox{Figure \ref{fig:timing}}. The horizontal axis represents the combined number of $P$ and $\pdot$ search points since the timing is only dependent on the total number. It does not depend differently on the number of $P$ versus $\pdot$ values assuming the $P$ dimension is much larger than the $\pdot$ dimension, which should be true in all cases.  For more information on specifics of our implementation of the CE, please see our Github repository, \href{https://github.com/mikekatz04/gce/tree/c3495ed7c1316cc26bb7362417d95d7832671c6f}{github.com/mikekatz04/gce}.

\section{Results}\label{sec:results}

For both the \cosmic$\ $and Burdge catalogs, we performed the CE entropy calculation with zero $\pdot$ values, as well as groups of 32, 128, and 512 $\pdot$ values. This allowed us to compare how this new ability to analyze the secondary axis of $\pdot$s compares to the original performance with a singular $\pdot$, generally taken to be zero. For clarity, performance in the context of this work means efficiently determining binary parameters to an acceptable degree of accuracy ($\sim10\%$), allowing observers to quickly filter information for targeted follow-up operations. Additionally, we can test how the density of the $\pdot$ dimension affects our detection capabilities. The idea here is to balance the advantage of accessing the $\pdot$ dimension against its computational cost.

We design our search grid based on general observing settings meaning we want to examine DWD detection from within a larger search for generic periodic objects. For both catalogs, we used the same group of periods; however, we determine our period grid by working with the associated frequencies to ensure the fastest orbits maintain a higher density within the grid. We set our maximum frequency to the inverse of a 3 minute period (480 days$^{-1}$). Our minimum frequency is two times the inverse of our baseline. In this work, the baseline, $B$, is 10 years. Therefore, our minimum frequency is $2/B=5.47\times10^{-4}$ days$^{-1}$. We then linearly space frequencies between these two values stepping in sizes of $\Delta f$, which we set to $1/(3B) = 9.13\times10^{-5}$ days$^{-1}$. This establishes 5,183,995 frequencies/periods to search over. The oversampling factor of 3 was chosen in accordance with current ZTF search methods \citep{Coughlin2020ZTF}. A further discussion on the choice of  oversampling factor can be found in \citet{Coughlin2020ZTF}.

For the $\pdot$ dimension, we chose values that represent the two data sets we analyzed. The \cosmic$\ $data set has a $\pdot$ range from $\sim-2.13\times10^{-12}$ to $-3.17\times10^{-11}$, while the Burdge data set varies from $\sim-1.22\times10^{-12}$ to $-5.44\times10^{-11}$. To cover these ranges, we log-space a grid from $-1\times10^{-12}$ to $-1\times10^{-10}$. We log-space the $\pdot$ dimension to maintain resolution where the decay is slower while searching over a smaller total number of points. Linear spacing did not change our results much for the quickly decaying binaries; however, for slowly decaying binaries, we found the linear spacing required more than the desired resolution. The number of points in these $\pdot$ grids is set to 32, 128, or 512.

\subsection{Period Recovery}\label{sec:period_recovery}

As previously mentioned, the goal of this algorithm is to locate sources by finding them with an above-threshold significance and to determine parameters to an \emph{acceptable} degree of accuracy (within $\sim10\%$). For information on the expected accuracy of the CE technique, please see \citet{Graham2013}. Assuming the observations will be followed up given a satisfactory significance, the period only needs rough determination. However, there is also an element of visual confirmation to sources with significance above the threshold. This visual confirmation means examining the phase-folded light curve, given the parameters determined by the algorithm, to see if there is structure exhibited by the source. Additionally, in order to train algorithms to identify sources through techniques like machine learning, it is desired that these period-finding algorithms locate a period that is a harmonic of the true period. 

\begin{figure*}
\begin{center}
\includegraphics[scale=0.5]{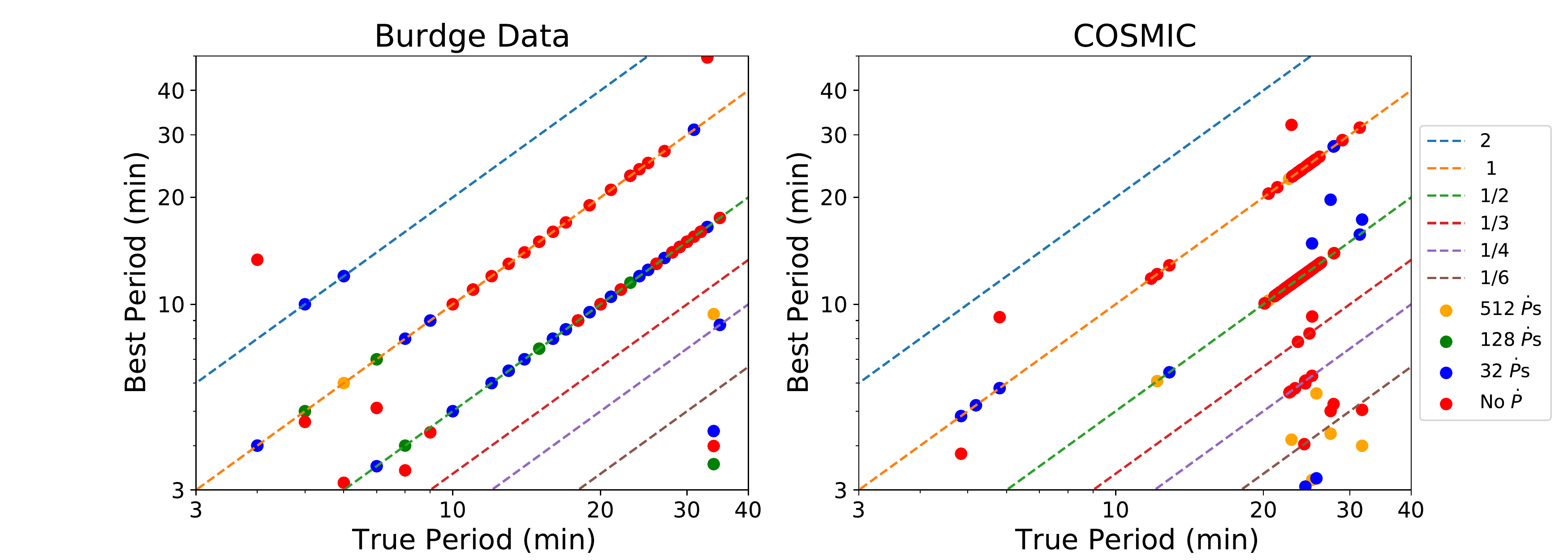}
\caption{A visualization of the period finding abilities of our CE implementation is shown above. The Burdge data set is shown on the left while the \cosmic$\ $data set is shown on the right. The horizontal axis represents the true period of a binary. The vertical axis represents the best period as found by the CE calculation. Therefore, points vertically separated represent the same binary light curve. Harmonics of the true period are expected when using period finding algorithms. For this reason, we have added lines representing 2, 1, 1/2, 1/3, 1/4, and 1/6 of the true period in blue, orange, green, red, purple, and brown dashed lines, respectively. Scatter points are plotted for searches with 512, 128, 32, and no $\pdot$s in orange, green, blue and red, respectively. The points are also layered in this order with the zero $\pdot$ search on top and 512 on the bottom. This layering is specifically chosen to illustrate at what search density is needed to properly constrain the detected period. For example, if only a red dot is seen for a specific binary (no other dots appear vertically above or below this dot), this means all $\pdot$ axis settings produced the same observed period. If a green dot is seen, this means the 128 $\pdot$ search returned a different period from the zero and 32 $\pdot$ searches. Assuming this observed green point is at a proper value for the observed period, this would indicate 128 $\pdot$s are generally needed to locate the proper period.}\label{fig:true_vs_best}
\end{center}
\end{figure*}

Since this work is based on the new capabilities of the algorithm, rather than the success rate of the CE method itself, we focus on the ability of the enhanced $\pdot$ search to help us detect more binaries in general, as well as extract parameters more accurately. \mbox{Figure \ref{fig:true_vs_best}} shows how the recovered periods compare to the true periods both with and without the $\pdot$ search. Points located directly above each other are from the same source. Additionally, the scatter points are layered by color purposefully to visually indicate the density of $\pdot$ sampling needed to better constrain binary parameters. The layer order is 512, 128, 32 and zero $\pdot$s with zero $\pdot$s on top. An example of the effect of this layering is if a point found testing with 32 $\pdot$s is plotted over a point found with 512, the 32 test will be the only one shown indicating that 32 $\pdot$s is adequate for locating the best parameters. For the Burdge data, it can be seen that including $\pdot$ better constrains the determined periods to period harmonics. This is especially true at fast periods where the one-dimensional search algorithm is unable to locate a harmonic period. Additionally, at these fast periods, a search with 32 $\pdot$ values is sufficient enough to locate harmonics rendering a 128 or 512 $\pdot$ search computationally unnecessary under these conditions.  At longer periods, the search without $\pdot$ inclusion is able to generally locate harmonics. Additionally, when using a $\pdot$ axis, the 32 $\pdot$ search is adequate for locating the optimal values at these longer periods. At the longest periods, there is some scatter away from the close harmonics due to lower significance observations.

A similar story is true for the \cosmic$\ $data set. Shorter periods require the $\pdot$ search to correctly identify a close harmonic of the true period. For longer periods, the period-only search is able to locate most periods at a close harmonic. Once again, at longer periods, lower significance observations scatter results away from the close harmonics.

\begin{figure*}
\begin{center}
\includegraphics[scale=0.4]{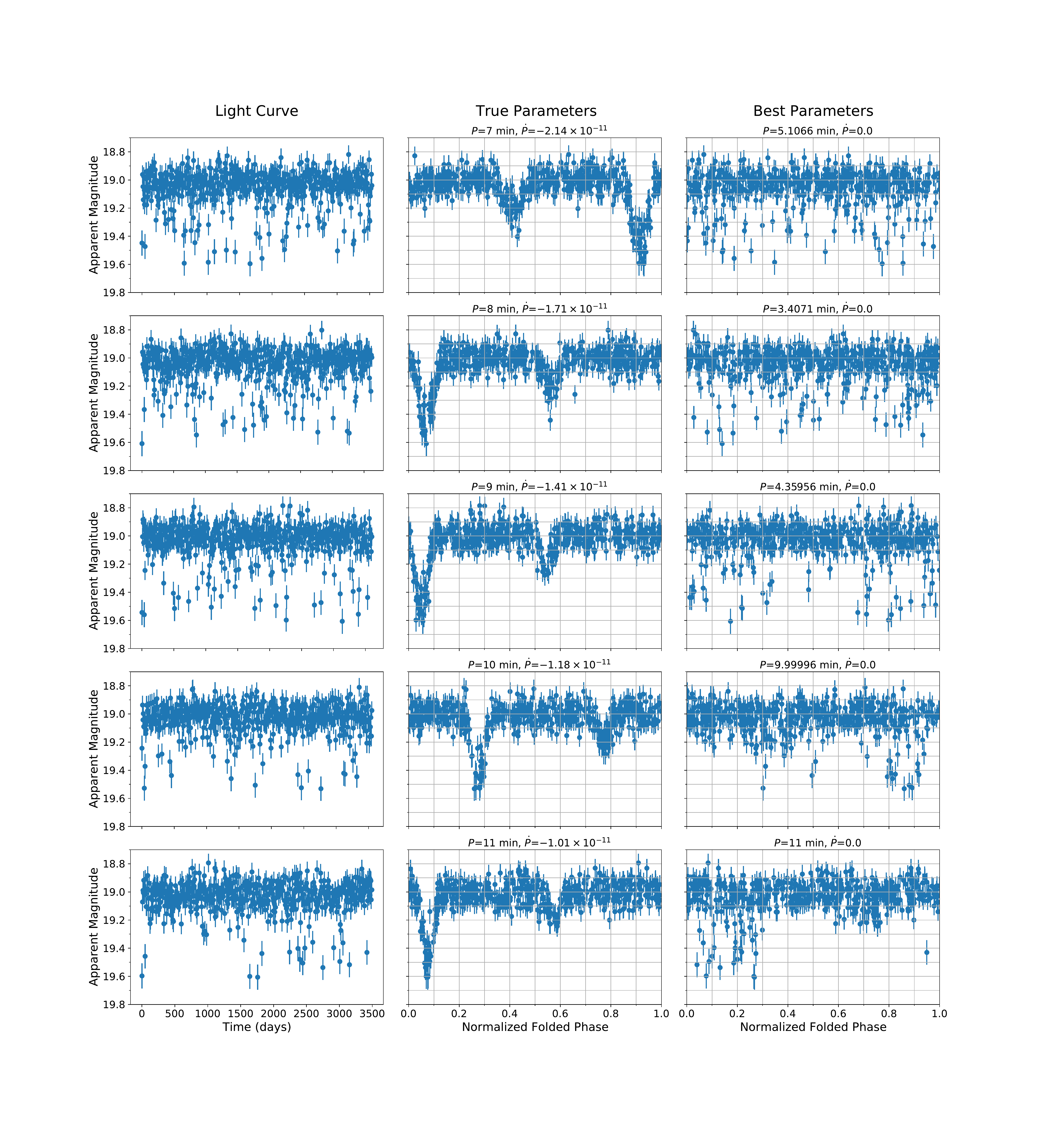}
\caption{Visual representations of the zero $\pdot$ search are shown above for 5 light curves from our Burdge set. The periods shown are 7, 8, 9, 10, and 11 min from top to bottom. Similar to \mbox{Figure \ref{fig:lc_ex}}, we show the observed light curve in the left column. The proper phase fold with the true period and $\pdot$ for each binary is shown in the center column. The right column shows the phase-folding using the best period found by the CE algorithm with $\pdot=0.0$ (this search also assumed $\pdot=0.0$). The true parameters and determined parameters are shown above each plot in the right two columns. It is clear visual follow up on a zero $\pdot$ search yields no consistent structure below $\sim9-10$ min. Please note these light curves are shown at an apparent magnitude of 19\,mag rather than the tested value of 16\,mag. At 16\,mag, the magnitude errors are small, making visual follow up simple. At 19 we can show a more realistic example of where the visual follow up highly depends on search results.}\label{fig:pdot_break}
\end{center}
\end{figure*}

Beyond just determining the best parameters, we want to ensure a detection of these sources via visual follow up. This scanning method is currently the popular technique in optical astronomy, including ZTF-based searches. Unfortunately, this makes understanding selection bias more difficult. A similar visual follow up technique is the ``waterfall plot'' used in pulsar searches \citep[e.g.][]{Amiri:2019bjk}. In order to ensure visual follow up is possible, we phase-fold the light curve based on the best values attained from the CE search. The goal here is to understand at what period the visual follow up can be completed without inclusion of the $\pdot$ axis during search. \mbox{Figure \ref{fig:pdot_break}} shows the phase folding of light curves progressing from a period of 7 min to 11 min.\footnote{The light curves are shown at an apparent magnitude of 19\,mag rather than the tested magnitude of 16\,mag. 16\,mag was necessary to ensure longer periods exhibit a detectable light curve. However, at $m\sim16$, the errors are small making visual follow up much easier. Therefore, we chose to visualize these light curves at $m=19$ to show a more realistic example where this $\pdot$ search makes a major difference in observing possible structure in the phase-folded light curve.} The central axis of plots shows folding with the true $P$ and $\pdot$; the right axis shows the folding with the best period found with the zero $\pdot$ search, as well as with $\pdot=0.0$. At a period of 7 min, The $\pdot=0.0$ light curve does not exhibit any structure indicating the correct period has been found. The same is true at 8 min. However, at 9 min, we begin to see the development of structure. At 10 min and 11 min, the fold is acceptable in terms of indicating there is some underlying structure. Therefore, under these settings, the rough turnover of the period where visual follow is possible without a $\pdot$ search is $\sim10$ min.

\subsection{Significance}\label{sec:res_sig}

Beyond locating an accurate parameterization of the binary, we must examine the significance of the detection. When running this algorithm on a large data set, such as the observations of ZTF or VRO, we must rely on its determination to provide possible follow up candidates. We assume that a significance of 10 is a dependable detection, consistent with ongoing searches \citep{burdge7min,coughlin40min}. Below that, especially around 5 or less, the detection is suspect. The CE could find a period that is near the true values. Conversely, completely incorrect parameters may be found when examining an actual periodic source due to signal aliasing or simply photometric noise. Additionally, it is possible for a non-periodic source to exhibit a significance value of 5 based on its light curve shape and associated errors. Therefore, we must be careful when designing automated algorithms because we cannot visually follow up on every trigger at a significance of 5.

\begin{figure}[h]
\begin{center}
\includegraphics[scale=0.32]{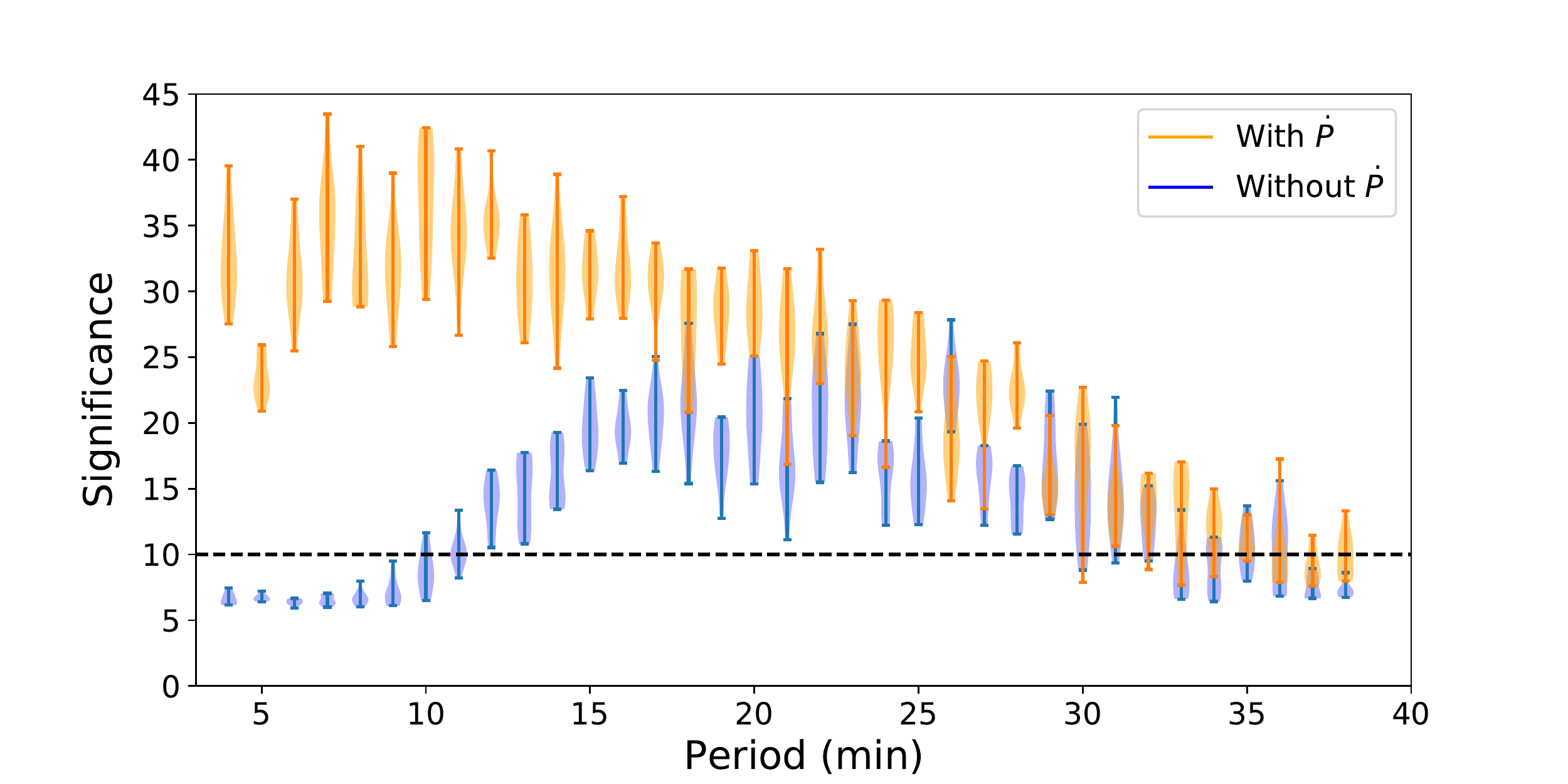}
\caption{A violin plot showing the significance comparing a $\pdot$ search to a search without considering a $\pdot$ dimension is shown above for our Burdge data set. Each set represents 10 randomly generated light curves using the same set of parameters. Blue points show the results of the search without a $\pdot$ axis. The search with a $\pdot$ dimension is shown in orange and represents a search over 32 $\pdot$ values. The significance is given by \mbox{Equation \ref{eq:signif}}. There is a black dashed line indicating a significance of 10, which represents our cut required for an automated detection.}\label{fig:sig_diff}
\end{center}
\end{figure}

With this in mind, we analyze the significance difference as $\pdot$ values are included in the CE search. \mbox{Figure \ref{fig:sig_diff}} shows the difference in the significance values obtained using a 32 $\pdot$ search versus a search without the $\pdot$ dimension for the Burdge data set. Both groups are represented with violin plots to show the distribution of the significance values obtained from the CE search at each period. In order to account for the random nature of errors when generating the light curves, we produced 10 sets of light curves with identical intrinsic magnitude, but unique random realizations of the observed light curves. 

It is clear from this plot the importance of searching over $\pdot$ values for the fastest binaries ($\lessapprox$ 10 min). Not only is the spread between the two cases considerable at a significance difference of $>25$, but the search without a $\pdot$ axis would not detect these sources in an automated framework. On the contrary, these sources would be easily detected with the inclusion of the $\pdot$ search. 

For periods between 10 and 15 min, two important changes occur. First, the spread in the significance values decreases. Over this range the significance observed with a $\pdot$ search decreases due to the changing morphology of the signal as the eclipse depth decreases at larger periods and more distant separations. Conversely, the significance of observations without the $\pdot$ dimension increases as the effect of the $\pdot$ over the observation time has a smaller impact on the dephasing of the system. The other important difference is that in this period range, the significance of observations without the $\pdot$ dimension are above the detectable threshold of 10. This changeover matches what we find in the visual turnover discussed in \mbox{Section \ref{sec:period_recovery}}. 

At longer periods, the significance of both methods of observation effectively decrease together as the eclipse depth decreases more and more. However, generally, we can expect the $\pdot$ search to return a slightly higher significance than without a $\pdot$ axis. This can be simply explained since the $\pdot$ search adds a degree of freedom allowing for more opportunity during phase folding to find a lower entropy. Interestingly, at the longest periods we tested, where the eclipse depth causes more difficulty for detection, some binaries can be automatically found with a $\pdot$ search, but not without (with a threshold significance of 10). This is not true within every random light curve realization, but, on average, this is the case.

\begin{figure}[h]
\begin{center}
\includegraphics[scale=0.32]{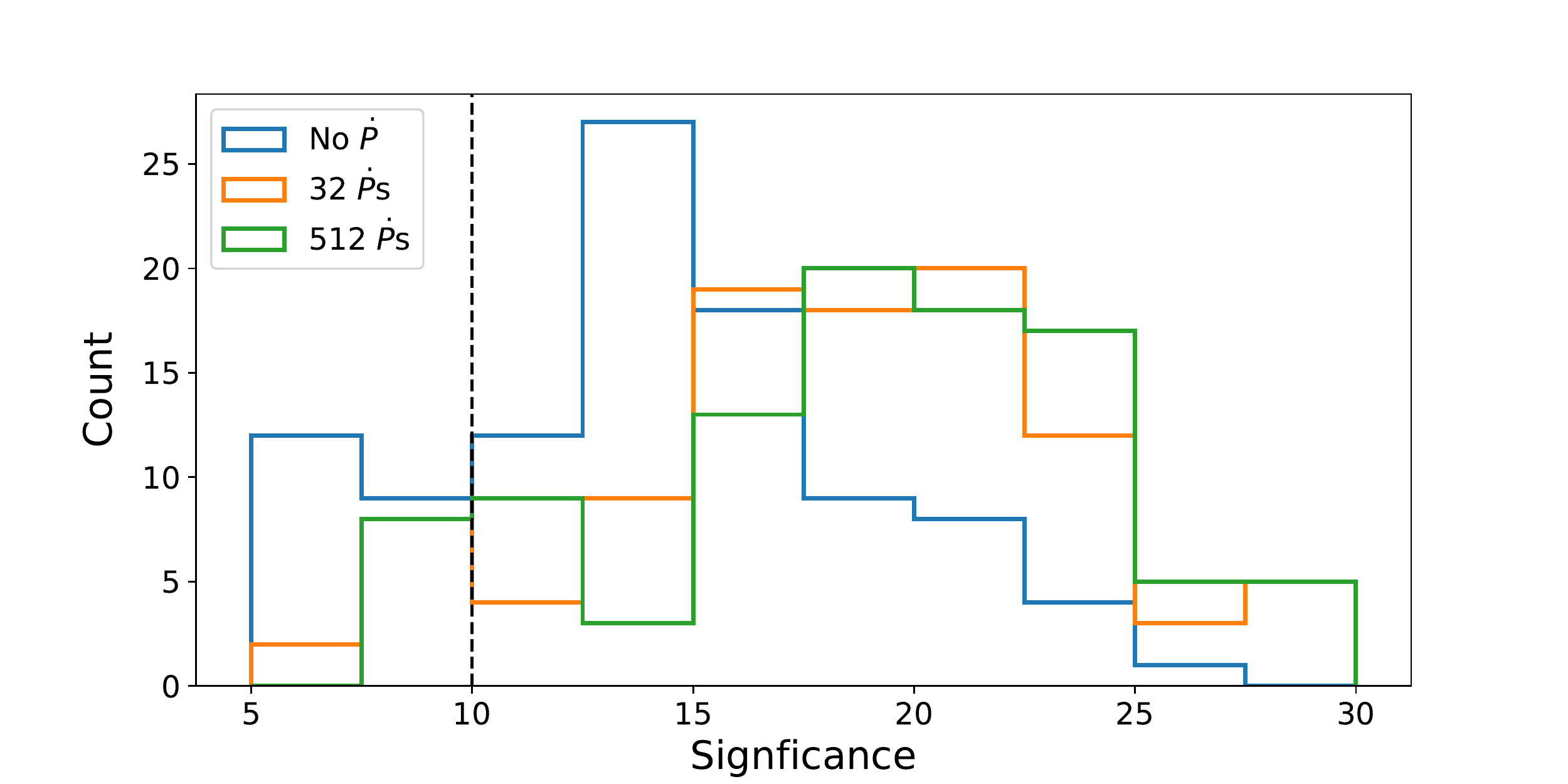}
\caption{Significance histograms for the \cosmic$\ $data set are shown above. The blue histogram represents the significance values returned by the search without a $\pdot$ axis. $\pdot$ searches with 32 and 512 $\pdot$ values are shown in orange and green, respectively. The significance is given in \mbox{Equation \ref{eq:signif}}. A vertical dashed line is added at a significance of 10 to indicate our detection threshold.}\label{fig:cosmic_sig_histogram}
\end{center}
\end{figure}

Due to the overlap in the periods from the \cosmic$\ $data, we represent the significance of observations from this set as histograms shown in \mbox{Figure \ref{fig:cosmic_sig_histogram}}. Since this data set has longer periods, we see similar behavior to the longer periods in \mbox{Figure \ref{fig:sig_diff}}. However, it is modulated slightly by the varying, rather than fixed, intrinsic parameters of the set. First, the $\pdot$ search, on average, returns higher significance values. However, the difference is small as these histograms overlap with the no $\pdot$ search results. Second, a similar behavior on the detectable threshold is seen. Based on the noise realization analyzed, 10 light curves in the \cosmic$\ $set are undetectable when employing a 32 $\pdot$ axis during search (8 are undetected with 512 $\pdot$s). On the contrary, 21 sources do not reach the detectable threshold when searching without a $\pdot$. This means, when running an automated search, 11 follow up indicators would be gained for sources of interest compared to a zero $\pdot$ search.

\section{Discussion}\label{sec:discuss}

The goal of an algorithm like CE is to iterate through large data sets as fast as possible and automatically highlight sources of interest. Additionally, for highlighted sources, the algorithm is expected to return a reasonable set of parameters describing the binary. Our new implementation of the CE improves upon both of these aspects. First, the algorithm is more efficient than previous versions allowing it to iterate through data sets more quickly. It also uses memory more efficiently.  Similarly, this expanded capability has led to an additional dimension, $\pdot$, over which to search. This provides more and stronger detections, and more precise parameterizations of observing targets.

The question then becomes balancing the computational cost of adding the $\pdot$ dimension to the search versus running over more sources without it. It is clear from our results (see \mbox{Section \ref{sec:results}}) that it is necessary to analyze this with different groups of periods in mind. To search for fast periods ($\lessapprox10$ min) where the $\pdot$ is significant enough to affect the phasing of a light curve over a long observation period, the $\pdot$ search must be incorporated. Without it, observations of fast binaries will not reach a detectable threshold in an automated sense, therefore, risking missing these unique binaries in a large catalog of sources. With that said, when focusing on these periods, it may be advantageous to tune the $\pdot$ axis towards the $\pdot$ values of interest to limit overhead caused by unlikely parameters. However, it must be kept in mind that this study has assumed General Relativistic $\pdot$ values that generally, but not fully, describe a variety of systems. For example, the 7 minute binary from \citet{burdge7min} exhibits $\pdot\approx-2.373\times10^{-11}$ss$^{-1}$, which is slightly higher in magnitude compared to the value predicted by General Relativity ($\pdot_\text{GW}=-2.184\times10^{-11}$ss$^{-1}$) due to the presence of tides \citep{burdge7min}. Therefore, in cases like this, the turnover in period-space where the $\pdot$ search is necessary will be at slightly longer periods. We must also note there is nothing in the CE algorithm specific to General Relativistic orbital decay. Over the timescales we are analyzing where we are only concerned with the factors changing the period to first order (see \mbox{Equation \ref{eq:P(t)}}), the generic $\pdot$ search we have used here would have similarly recovered light curves from binaries exhibiting tidal corrections to the orbital decay. We have verified this by successfully recovering the true $\pdot$ for the 7 min binary using actual data, including its contribution from tides. Similar to orbital decay rates that differ from $\pdot_\text{GW}$, we have verified our algorithm works for $\pdot>0.0$, as is expected, assuming the $\pdot$ axis search values are adjusted accordingly. Due to the general method for phase folding (see \mbox{Equation \ref{eq:phase_fold}}), the internals of the code do not change when searching positive $\pdot$ values. Therefore, when searching in large surveys for DWD binary sources with short periods, it may be advantageous to search over positive and negative $\pdot$ values simultaneously.

When considering longer periods, depending on the application, it may not be advantageous to search over $\pdot$. It may be beneficial, especially in larger catalogs, to search without a $\pdot$ to narrow sources to those of interest based on a more lenient significance limit. Then, with this confined catalog, perform the $\pdot$ search in order to ensure the best detection capabilities. This is dependent on the size of the catalog and the exact sources of interest. Sources with longer periods than those tested here are likely not to benefit from any search over $\pdot$ values.  

One additional aspect that must be considered when assessing the computational cost of the $\pdot$ search is the memory necessary to perform this search. Specifically, the algorithm must store in memory all CE values at each grid point. Therefore, with a large and dense grid of period values, searching over $\pdot$ values will linearly increase this required memory. This memory limit is dependent on the number of light curves, $\pdot$s, and $P$s chosen to iterate over, as well as the specific hardware used. This is a balancing act that must be tuned to each case usage and GPU device. As previously discussed in \mbox{Section \ref{sec:gpu_impl}}, we have added to the algorithm the ability to batch groups of light curves and $\pdot$ values. This performs the calculation on a batch, locates the minimum CE and its associated parameters, and stores these values. After iterating through all batches, these minimum values from each batch are then compared against one another to find the global minimum. Additionally, we carefully combine the statistics of the CE values from each batch to ensure our significance calculation is legitimate.

Since our algorithm is only a more efficient version of the CE, the code produced by this work can be used to search for all periodic sources. We have not tested other sources in this work as we have focused on sources where the $\pdot$ will factor into the analysis. However, as this algorithm is faster and more memory efficient than previous iterations, we recommend its usage in any situation where the CE calculation is employed.

Algorithms like \texttt{gce} will be crucial in searching real world data sets such as those produced by VRO and ZTF. Acceleration searches will be particularly crucial for finding any system undergoing significant orbital decay in VRO data. This data will be sparsley sampled over the course of a decade \citep{2019LSST}, such that most of the strongest LISA counterparts in its data set will not maintain coherence over the observation timescale, requiring a sweep of both period and $\pdot$ to identify such systems. In more densley sampled surveys like the Zwicky Transient Facility, strong LISA sources such as ZTF J1539+5027 have already been identified; however, an acceleration search could be particularly useful as a tool for identifying certain classes of systems such as eclipsing double CO WD pairs. CO WD binaries have very narrow eclipses, and thus decohere more quickly than their counterparts containing He WDs, which exhibit broader eclipses and slower orbital decay at the same period.

\section{Conclusion}\label{sec:conc}

We present a new, GPU-accelerated implementation of the Conditional Entropy, a popular periodic source detection algorithm from \cite{Graham2013}. Our implementation is $\sim5$x faster than the previous implementation. It scales well when searching over a larger grid of search parameters, as well as at a variety of input light curve lengths and numbers of phase bins (see \mbox{Figure \ref{fig:timing}}). Our code is called \texttt{gce} and is publicly available for general usage at \href{https://github.com/mikekatz04/gce/tree/c3495ed7c1316cc26bb7362417d95d7832671c6f}{github.com/mikekatz04/gce}.

This improved efficiency has allowed us to expand the one-dimensional search grid over the period, $P$, to two dimensions with the addition of the time derivative of the period, $\pdot$. Assuming the period grid is sufficiently dense, the addition of the $\pdot$ search affects the algorithm duration by linearly scaling with the number of $\pdot$s. 

We compared the one-dimensional search to a two-dimensional search by examining the ability of each to find the true period, as well as the differences in the significance associated with the minimum CE value. To perform this analysis, we analyzed two data sets. The first data set was chosen to analyze periods and their associated gravitational radiation-generated orbital decay on a grid with fixed intrinsic parameters. These parameters were chosen to be the same as those measured for the 7 min binary from \citet{burdge7min}. The second data set was a small hand-picked subset of a realistic population of DWD binaries created using the \cosmic$\ $population synthesis code \citep{COSMICcode, Breivik2019}. With steps described in \mbox{Section \ref{sec:genpop}}, we narrowed this data set to 100 binaries. 

We found that including the $\pdot$ search is absolutely necessary for binaries at periods less than $\sim10$ min. For this group of binaries, we could only analyze our Burdge data set. In terms of accurate determination of periods, a search without $\pdot$s will produce sporadic results determining periods farther from the true value at low significance. Employing a $\pdot$ axis in the search returns periods that are effectively equivalent to the true value or a close harmonic as is expected with period finding algorithms. These observations including a $\pdot$ axis were found with much higher significance compared to the $\pdot=0.0$ case. This difference reached $\gtrapprox25$. Additionally, sources with these short periods are undetectable if we require a significance of 10 for detection. Therefore, including $\pdot$ in the search is necessary for detecting these sources in an automated setting. Similarly, we find when we examine the folded light curves visually, we see that up to $\sim10$ min, the best found period with a $\pdot=0.0$ search produces a visually erratic phase folding not providing any indication to the observer that the correct period has been found. 

At longer periods above ten minutes, sources are detectable (significance of 10 or more) even when searching without a $\pdot$ dimension. For all observations, it can easily be seen that including $\pdot$ will raise the significance even if a binary is already detectable just based on searching over an additional degree of freedom in terms of phase-folding light curves. This is shown using both our Burdge and \cosmic$\ $data sets. Therefore, for binaries on the fringe of detection, adding $\pdot$ to the search may help salvage some lost sources. For these longer periods, the algorithm constrains the period well even without a $\pdot$ dimension.

From these findings, we discuss real-world observing strategies and consequences in \mbox{Section \ref{sec:discuss}}. In general, the main question is balancing the computational cost of the $\pdot$ search with its advantages of locating more sources, as well as providing better parameterization of these sources.

Ultimately, this more efficient algorithm provides new avenues for analysis by allowing for more accurate searches, as well as searching more data at a higher rate. This quality will become increasingly more important as massive amounts of data are collected with larger and more expansive surveys. 

\section*{Data Availability}

The data underlying this article are publicly available in the \texttt{gce} code repository on \href{https://github.com/mikekatz04/gce/tree/c3495ed7c1316cc26bb7362417d95d7832671c6f}{Github}.

\section*{Acknowledgements}
M.L.K. acknowledges support from the National Science Foundation under grant DGE-0948017 and the Chateaubriand Fellowship from the Office for Science \& Technology of the Embassy of France in the United States. O.R.C. gratefully acknowledges support from the LIGO Scientific Collaboration, the California Institute of Technology, and the National Science Foundation through the LIGO Summer Undergraduate Research Fellowships (LIGO SURF) Program hosted by Caltech Student-Faculty Programs.
M.~W.~Coughlin acknowledges support from the National Science Foundation with grant number PHY-2010970.
This research was supported in part through the computational resources and staff contributions provided for the Quest/Grail high performance computing facility at Northwestern University. Astropy, a community-developed core Python package for Astronomy, was used in this research \citep{Astropy}. This paper also employed use of Scipy \citep{scipy}, Numpy \citep{Numpy}, Pandas \citep{reback2020pandas, mckinney-proc-scipy-2010}, and Matplotlib \citep{Matplotlib}.

\bibliographystyle{aasjournal}

\begin{thebibliography}{}
\expandafter\ifx\csname natexlab\endcsname\relax\def\natexlab#1{#1}\fi
\providecommand{\url}[1]{\href{#1}{#1}}
\providecommand{\dodoi}[1]{doi:~\href{http://doi.org/#1}{\nolinkurl{#1}}}
\providecommand{\doeprint}[1]{\href{http://ascl.net/#1}{\nolinkurl{http://ascl.net/#1}}}
\providecommand{\doarXiv}[1]{\href{https://arxiv.org/abs/#1}{\nolinkurl{https://arxiv.org/abs/#1}}}

\bibitem[{{Amaro-Seoane} {et~al.}(2017){Amaro-Seoane}, {Audley}, {Babak},
  {Baker}, {Barausse}, {Bender}, {Berti}, {Binetruy}, {Born}, {Bortoluzzi},
  {Camp}, {Caprini}, {Cardoso}, {Colpi}, {Conklin}, {Cornish}, {Cutler},
  {Danzmann}, {Dolesi}, {Ferraioli}, {Ferroni}, {Fitzsimons}, {Gair}, {Gesa
  Bote}, {Giardini}, {Gibert}, {Grimani}, {Halloin}, {Heinzel}, {Hertog},
  {Hewitson}, {Holley-Bockelmann}, {Hollington}, {Hueller}, {Inchauspe},
  {Jetzer}, {Karnesis}, {Killow}, {Klein}, {Klipstein}, {Korsakova}, {Larson},
  {Livas}, {Lloro}, {Man}, {Mance}, {Martino}, {Mateos}, {McKenzie},
  {McWilliams}, {Miller}, {Mueller}, {Nardini}, {Nelemans}, {Nofrarias},
  {Petiteau}, {Pivato}, {Plagnol}, {Porter}, {Reiche}, {Robertson},
  {Robertson}, {Rossi}, {Russano}, {Schutz}, {Sesana}, {Shoemaker}, {Slutsky},
  {Sopuerta}, {Sumner}, {Tamanini}, {Thorpe}, {Troebs}, {Vallisneri},
  {Vecchio}, {Vetrugno}, {Vitale}, {Volonteri}, {Wanner}, {Ward}, {Wass},
  {Weber}, {Ziemer}, \& {Zweifel}}]{LISA2017}
{Amaro-Seoane}, P., {Audley}, H., {Babak}, S., {et~al.} 2017, arXiv e-prints,
  arXiv:1702.00786.
\newblock \doarXiv{1702.00786}

\bibitem[{Amiri {et~al.}(2019)}]{Amiri:2019bjk}
Amiri, M., {et~al.} 2019, Nature, 566, 235, \dodoi{10.1038/s41586-018-0864-x}

\bibitem[{{Astropy Collaboration} {et~al.}(2013){Astropy Collaboration},
  Robitaille, Tollerud, Greenfield, Droettboom, Bray, Aldcroft, Davis,
  Ginsburg, Price-Whelan, Kerzendorf, Conley, Crighton, Barbary, Muna,
  Ferguson, Grollier, Parikh, Nair, Unther, Deil, Woillez, Conseil, Kramer,
  Turner, Singer, Fox, Weaver, Zabalza, Edwards, Azalee~Bostroem, Burke, Casey,
  Crawford, Dencheva, Ely, Jenness, Labrie, Lim, Pierfederici, Pontzen, Ptak,
  Refsdal, Servillat, \& Streicher}]{Astropy}
{Astropy Collaboration}, Robitaille, T.~P., Tollerud, E.~J., {et~al.} 2013,
  \aap, 558, A33, \dodoi{10.1051/0004-6361/201322068}

\bibitem[{Behnel {et~al.}(2011)Behnel, Bradshaw, Citro, Dalcin, Seljebotn, \&
  Smith}]{Cython}
Behnel, S., Bradshaw, R., Citro, C., {et~al.} 2011, Computing in Science \&
  Engineering, 13, 31

\bibitem[{{Bellm} {et~al.}(2019){Bellm}, {Kulkarni}, {Graham}, {Dekany},
  {Smith}, {Riddle}, {Masci}, {Helou}, {Prince}, {Adams}, {Barbarino},
  {Barlow}, {Bauer}, {Beck}, {Belicki}, {Biswas}, {Blagorodnova}, {Bodewits},
  {Bolin}, {Brinnel}, {Brooke}, {Bue}, {Bulla}, {Burruss}, {Cenko}, {Chang},
  {Connolly}, {Coughlin}, {Cromer}, {Cunningham}, {De}, {Delacroix}, {Desai},
  {Duev}, {Eadie}, {Farnham}, {Feeney}, {Feindt}, {Flynn}, {Franckowiak},
  {Frederick}, {Fremling}, {Gal-Yam}, {Gezari}, {Giomi}, {Goldstein},
  {Golkhou}, {Goobar}, {Groom}, {Hacopians}, {Hale}, {Henning}, {Ho}, {Hover},
  {Howell}, {Hung}, {Huppenkothen}, {Imel}, {Ip}, {Ivezi{\'c}}, {Jackson},
  {Jones}, {Juric}, {Kasliwal}, {Kaspi}, {Kaye}, {Kelley}, {Kowalski},
  {Kramer}, {Kupfer}, {Landry}, {Laher}, {Lee}, {Lin}, {Lin}, {Lunnan},
  {Giomi}, {Mahabal}, {Mao}, {Miller}, {Monkewitz}, {Murphy}, {Ngeow},
  {Nordin}, {Nugent}, {Ofek}, {Patterson}, {Penprase}, {Porter}, {Rauch},
  {Rebbapragada}, {Reiley}, {Rigault}, {Rodriguez}, {van Roestel}, {Rusholme},
  {van Santen}, {Schulze}, {Shupe}, {Singer}, {Soumagnac}, {Stein}, {Surace},
  {Sollerman}, {Szkody}, {Taddia}, {Terek}, {Van Sistine}, {van Velzen},
  {Vestrand}, {Walters}, {Ward}, {Ye}, {Yu}, {Yan}, \&
  {Zolkower}}]{2019ZTFsystemoverview}
{Bellm}, E.~C., {Kulkarni}, S.~R., {Graham}, M.~J., {et~al.} 2019, \pasp, 131,
  018002, \dodoi{10.1088/1538-3873/aaecbe}

\bibitem[{{Boehle} {et~al.}(2016){Boehle}, {Ghez}, {Sch{\"o}del}, {Meyer},
  {Yelda}, {Albers}, {Martinez}, {Becklin}, {Do}, {Lu}, {Matthews}, {Morris},
  {Sitarski}, \& {Witzel}}]{Boehle2016SagADistance}
{Boehle}, A., {Ghez}, A.~M., {Sch{\"o}del}, R., {et~al.} 2016, \apj, 830, 17,
  \dodoi{10.3847/0004-637X/830/1/17}

\bibitem[{{Breivik} {et~al.}(2019{\natexlab{a}}){Breivik}, {Coughlin}, {Zevin},
  {Rodriguez}, {Andrews}, {Kimball}, {mcdigman}, \& {1nhtran}}]{COSMICcode}
{Breivik}, K., {Coughlin}, S., {Zevin}, M., {et~al.} 2019{\natexlab{a}},
  {COSMIC-PopSynth/COSMIC: COSMIC}, v3.2.0,  Zenodo,
  \dodoi{10.5281/zenodo.3561144}

\bibitem[{{Breivik} {et~al.}(2018){Breivik}, {Kremer}, {Bueno}, {Larson},
  {Coughlin}, \& {Kalogera}}]{Breivik2018}
{Breivik}, K., {Kremer}, K., {Bueno}, M., {et~al.} 2018, \apjl, 854, L1,
  \dodoi{10.3847/2041-8213/aaaa23}

\bibitem[{{Breivik} {et~al.}(2019{\natexlab{b}}){Breivik}, {Coughlin}, {Zevin},
  {Rodriguez}, {Kremer}, {Ye}, {Andrews}, {Kurkowski}, {Digman}, {Larson}, \&
  {Rasio}}]{Breivik2019}
{Breivik}, K., {Coughlin}, S.~C., {Zevin}, M., {et~al.} 2019{\natexlab{b}},
  arXiv e-prints, arXiv:1911.00903.
\newblock \doarXiv{1911.00903}

\bibitem[{Brough {et~al.}(2020)Brough, Collins, Demarco, Ferguson, Galaz,
  Holwerda, Martinez-Lombilla, Mihos, \& Montes}]{VRO2020}
Brough, S., Collins, C., Demarco, R., {et~al.} 2020, arXiv e-prints,
  arXiv:2001.11067

\bibitem[{{Burdge} {et~al.}(2019){Burdge}, {Coughlin}, {Fuller}, {Kupfer},
  {Bellm}, {Bildsten}, {Graham}, {Kaplan}, {van Roestel}, {Dekany}, {Duev},
  {Feeney}, {Giomi}, {Helou}, {Kaye}, {Laher}, {Mahabal}, {Masci}, {Riddle},
  {Shupe}, {Soumagnac}, {Smith}, {Szkody}, {Walters}, {Kulkarni}, \&
  {Prince}}]{burdge7min}
{Burdge}, K.~B., {Coughlin}, M.~W., {Fuller}, J., {et~al.} 2019, \nat, 571,
  528, \dodoi{10.1038/s41586-019-1403-0.}

\bibitem[{{Coughlin} {et~al.}(2019){Coughlin}, {Dekany}, {Duev}, {Feeney},
  {Kulkarni}, {Riddle}, {Ahumada}, {Burdge}, {Dugas}, {Fremling}, {Hallinan},
  {Prince}, \& {van Roestel}}]{KPED2019}
{Coughlin}, M.~W., {Dekany}, R.~G., {Duev}, D.~A., {et~al.} 2019, \mnras, 485,
  1412, \dodoi{10.1093/mnras/stz497}

\bibitem[{Coughlin {et~al.}(2020)Coughlin, Burdge, Sterl Phinney, van Roestel,
  Bellm, Dekany, Delacroix, Duev, Feeney, Graham, \& et~al.}]{coughlin40min}
Coughlin, M.~W., Burdge, K., Sterl Phinney, E., {et~al.} 2020, MNRAS: Letters,
  494, L91–L96, \dodoi{10.1093/mnrasl/slaa044}

\bibitem[{{Coughlin} {et~al.}(2020){Coughlin}, {Burdge}, {Duev}, {Katz}, {van
  Roestel}, {Drake}, {Graham}, {Hillenbrand}, {Mahabal}, {Masci}, {Mr{\'o}z},
  {Prince}, {Yao}, {Bellm}, {Burruss}, {Dekany}, {Jaodand}, {Kaplan}, {Kupfer},
  {Laher}, {Riddle}, {Rigault}, {Rodriguez}, {Rusholme}, \&
  {Zolkower}}]{Coughlin2020ZTF}
{Coughlin}, M.~W., {Burdge}, K., {Duev}, D.~A., {et~al.} 2020, arXiv e-prints,
  arXiv:2009.14071.
\newblock \doarXiv{2009.14071}

\bibitem[{{Graham} {et~al.}(2013){Graham}, {Drake}, {Djorgovski}, {Mahabal}, \&
  {Donalek}}]{Graham2013}
{Graham}, M.~J., {Drake}, A.~J., {Djorgovski}, S.~G., {Mahabal}, A.~A., \&
  {Donalek}, C. 2013, \mnras, 434, 2629, \dodoi{10.1093/mnras/stt1206}

\bibitem[{Graham {et~al.}(2013)Graham, Drake, Djorgovski, Mahabal, Donalek,
  Duan, \& Maker}]{Graham2013period_finding}
Graham, M.~J., Drake, A.~J., Djorgovski, S.~G., {et~al.} 2013, \mnras, 434,
  3423, \dodoi{10.1093/mnras/stt1264}

\bibitem[{{Graham} {et~al.}(2019){Graham}, {Kulkarni}, {Bellm}, {Adams},
  {Barbarino}, {Blagorodnova}, {Bodewits}, {Bolin}, {Brady}, {Cenko}, {Chang},
  {Coughlin}, {De}, {Eadie}, {Farnham}, {Feindt}, {Franckowiak}, {Fremling},
  {Gezari}, {Ghosh}, {Goldstein}, {Golkhou}, {Goobar}, {Ho}, {Huppenkothen},
  {Ivezi{\'c}}, {Jones}, {Juric}, {Kaplan}, {Kasliwal}, {Kelley}, {Kupfer},
  {Lee}, {Lin}, {Lunnan}, {Mahabal}, {Miller}, {Ngeow}, {Nugent}, {Ofek},
  {Prince}, {Rauch}, {van Roestel}, {Schulze}, {Singer}, {Sollerman}, {Taddia},
  {Yan}, {Ye}, {Yu}, {Barlow}, {Bauer}, {Beck}, {Belicki}, {Biswas}, {Brinnel},
  {Brooke}, {Bue}, {Bulla}, {Burruss}, {Connolly}, {Cromer}, {Cunningham},
  {Dekany}, {Delacroix}, {Desai}, {Duev}, {Feeney}, {Flynn}, {Frederick},
  {Gal-Yam}, {Giomi}, {Groom}, {Hacopians}, {Hale}, {Helou}, {Henning},
  {Hover}, {Hillenbrand}, {Howell}, {Hung}, {Imel}, {Ip}, {Jackson}, {Kaspi},
  {Kaye}, {Kowalski}, {Kramer}, {Kuhn}, {Landry}, {Laher}, {Mao}, {Masci},
  {Monkewitz}, {Murphy}, {Nordin}, {Patterson}, {Penprase}, {Porter},
  {Rebbapragada}, {Reiley}, {Riddle}, {Rigault}, {Rodriguez}, {Rusholme}, {van
  Santen}, {Shupe}, {Smith}, {Soumagnac}, {Stein}, {Surace}, {Szkody}, {Terek},
  {Van Sistine}, {van Velzen}, {Vestrand}, {Walters}, {Ward}, {Zhang}, \&
  {Zolkower}}]{2019ZTFsciobj}
{Graham}, M.~J., {Kulkarni}, S.~R., {Bellm}, E.~C., {et~al.} 2019, \pasp, 131,
  078001, \dodoi{10.1088/1538-3873/ab006c}

\bibitem[{Harding {et~al.}(2016)Harding, Hallinan, Milburn, Gardner, Konidaris,
  Singh, Shao, Sandhu, Kyne, \& Schlichting}]{HaHa2016}
Harding, L.~K., Hallinan, G., Milburn, J., {et~al.} 2016, MNRAS, 457, 3036,
  \dodoi{10.1093/mnras/stw094}

\bibitem[{{Hoffman}(2017)}]{cuvarbase}
{Hoffman}, J. 2017, {cuvarbase}.
\newblock \url{https://github.com/johnh2o2/cuvarbase}

\bibitem[{Hunter(2007)}]{Matplotlib}
Hunter, J.~D. 2007, Computing In Science {\&} Engineering, 9, 90,
  \dodoi{10.1109/MCSE.2007.55}

\bibitem[{{Hurley} {et~al.}(2000){Hurley}, {Pols}, \&
  {Tout}}]{Hurley2000StellarEvo}
{Hurley}, J.~R., {Pols}, O.~R., \& {Tout}, C.~A. 2000, \mnras, 315, 543,
  \dodoi{10.1046/j.1365-8711.2000.03426.x}

\bibitem[{{Ivezi{\'c}} {et~al.}(2019){Ivezi{\'c}}, {Kahn}, {Tyson}, {Abel},
  {Acosta}, {Allsman}, {Alonso}, {AlSayyad}, {Anderson}, {Andrew}, \&
  et~al.}]{2019LSST}
{Ivezi{\'c}}, {\v{Z}}., {Kahn}, S.~M., {Tyson}, J.~A., {et~al.} 2019, \apj,
  873, 111, \dodoi{10.3847/1538-4357/ab042c}

\bibitem[{Johnston \& Karastergiou(2017)}]{Johnston:2017wgm}
Johnston, S., \& Karastergiou, A. 2017, Mon. Not. Roy. Astron. Soc., 467, 3493,
  \dodoi{10.1093/mnras/stx377}

\bibitem[{Jones {et~al.}(2018)Jones, Oliphant, Peterson, \& {others}}]{scipy}
Jones, E., Oliphant, T., Peterson, P., \& {others}. 2018, {Tesla V100
  Performance Guide}.
\newblock
  \url{https://www.nvidia.com/content/dam/en-zz/Solutions/Data-Center/tesla-product-literature/v100-application-performance-guide.pdf}

\bibitem[{{Korol} {et~al.}(2017){Korol}, {Rossi}, {Groot}, {Nelemans},
  {Toonen}, \& {Brown}}]{2017korol}
{Korol}, V., {Rossi}, E.~M., {Groot}, P.~J., {et~al.} 2017, \mnras, 470, 1894,
  \dodoi{10.1093/mnras/stx1285}

\bibitem[{Korol {et~al.}(2020)Korol, Toonen, Klein, Belokurov, Vincenzo,
  Buscicchio, Gerosa, Moore, Roebber, Rossi, \&
  Vecchio}]{Korol2020Populations_DWD}
Korol, V., Toonen, S., Klein, A., {et~al.} 2020, arXiv e-prints,
  arXiv:2002.10462

\bibitem[{{Kremer} {et~al.}(2015){Kremer}, {Sepinsky}, \&
  {Kalogera}}]{Kremer2015}
{Kremer}, K., {Sepinsky}, J., \& {Kalogera}, V. 2015, \apj, 806, 76,
  \dodoi{10.1088/0004-637X/806/1/76}

\bibitem[{{Kupfer} {et~al.}(2018){Kupfer}, {Korol}, {Shah}, {Nelemans},
  {Marsh}, {Ramsay}, {Groot}, {Steeghs}, \& {Rossi}}]{LISAbinaries2018}
{Kupfer}, T., {Korol}, V., {Shah}, S., {et~al.} 2018, \mnras, 480, 302,
  \dodoi{10.1093/mnras/sty1545}

\bibitem[{Lamberts {et~al.}(2019)Lamberts, Blunt, Littenberg, Garrison-Kimmel,
  Kupfer, \& Sanderson}]{Lamberts2019LISA_DWD}
Lamberts, A., Blunt, S., Littenberg, T.~B., {et~al.} 2019, \mnras, 490, 5888,
  \dodoi{10.1093/mnras/stz2834}

\bibitem[{Lau {et~al.}(2020)Lau, Mandel, Vigna-G{\'{o}}mez, Neijssel,
  Stevenson, \& Sesana}]{Lau2020BNS_LISA}
Lau, M. Y.~M., Mandel, I., Vigna-G{\'{o}}mez, A., {et~al.} 2020, \mnras, 492,
  3061, \dodoi{10.1093/mnras/staa002}

\bibitem[{{LSST Science Collaboration} {et~al.}(2009){LSST Science
  Collaboration}, {Abell}, {Allison}, {Anderson}, {Andrew}, {Angel}, {Armus},
  {Arnett}, {Asztalos}, {Axelrod}, {Bailey}, {Ballantyne}, {Bankert},
  {Barkhouse}, {Barr}, {Barrientos}, {Barth}, {Bartlett}, {Becker}, {Becla},
  {Beers}, {Bernstein}, {Biswas}, {Blanton}, {Bloom}, {Bochanski}, {Boeshaar},
  {Borne}, {Bradac}, {Brandt}, {Bridge}, {Brown}, {Brunner}, {Bullock},
  {Burgasser}, {Burge}, {Burke}, {Cargile}, {Chand rasekharan}, {Chartas},
  {Chesley}, {Chu}, {Cinabro}, {Claire}, {Claver}, {Clowe}, {Connolly}, {Cook},
  {Cooke}, {Cooray}, {Covey}, {Culliton}, {de Jong}, {de Vries}, {Debattista},
  {Delgado}, {Dell'Antonio}, {Dhital}, {Di Stefano}, {Dickinson}, {Dilday},
  {Djorgovski}, {Dobler}, {Donalek}, {Dubois-Felsmann}, {Durech},
  {Eliasdottir}, {Eracleous}, {Eyer}, {Falco}, {Fan}, {Fassnacht}, {Ferguson},
  {Fernandez}, {Fields}, {Finkbeiner}, {Figueroa}, {Fox}, {Francke}, {Frank},
  {Frieman}, {Fromenteau}, {Furqan}, {Galaz}, {Gal-Yam}, {Garnavich},
  {Gawiser}, {Geary}, {Gee}, {Gibson}, {Gilmore}, {Grace}, {Green}, {Gressler},
  {Grillmair}, {Habib}, {Haggerty}, {Hamuy}, {Harris}, {Hawley}, {Heavens},
  {Hebb}, {Henry}, {Hileman}, {Hilton}, {Hoadley}, {Holberg}, {Holman},
  {Howell}, {Infante}, {Ivezic}, {Jacoby}, {Jain}, {R}, {Jedicke}, {Jee},
  {Garrett Jernigan}, {Jha}, {Johnston}, {Jones}, {Juric}, {Kaasalainen},
  {Styliani}, {Kafka}, {Kahn}, {Kaib}, {Kalirai}, {Kantor}, {Kasliwal},
  {Keeton}, {Kessler}, {Knezevic}, {Kowalski}, {Krabbendam}, {Krughoff},
  {Kulkarni}, {Kuhlman}, {Lacy}, {Lepine}, {Liang}, {Lien}, {Lira}, {Long},
  {Lorenz}, {Lotz}, {Lupton}, {Lutz}, {Macri}, {Mahabal}, {Mandelbaum},
  {Marshall}, {May}, {McGehee}, {Meadows}, {Meert}, {Milani}, {Miller},
  {Miller}, {Mills}, {Minniti}, {Monet}, {Mukadam}, {Nakar}, {Neill}, {Newman},
  {Nikolaev}, {Nordby}, {O'Connor}, {Oguri}, {Oliver}, {Olivier}, {Olsen},
  {Olsen}, {Olszewski}, {Oluseyi}, {Padilla}, {Parker}, {Pepper}, {Peterson},
  {Petry}, {Pinto}, {Pizagno}, {Popescu}, {Prsa}, {Radcka}, {Raddick},
  {Rasmussen}, {Rau}, {Rho}, {Rhoads}, {Richards}, {Ridgway}, {Robertson},
  {Roskar}, {Saha}, {Sarajedini}, {Scannapieco}, {Schalk}, {Schindler},
  {Schmidt}, {Schmidt}, {Schneider}, {Schumacher}, {Scranton}, {Sebag},
  {Seppala}, {Shemmer}, {Simon}, {Sivertz}, {Smith}, {Allyn Smith}, {Smith},
  {Spitz}, {Stanford}, {Stassun}, {Strader}, {Strauss}, {Stubbs}, {Sweeney},
  {Szalay}, {Szkody}, {Takada}, {Thorman}, {Trilling}, {Trimble}, {Tyson}, {Van
  Berg}, {Vand en Berk}, {VanderPlas}, {Verde}, {Vrsnak}, {Walkowicz}, {Wand
  elt}, {Wang}, {Wang}, {Warner}, {Wechsler}, {West}, {Wiecha}, {Williams},
  {Willman}, {Wittman}, {Wolff}, {Wood-Vasey}, {Wozniak}, {Young}, {Zentner},
  \& {Zhan}}]{2009LSSTsciencebook}
{LSST Science Collaboration}, {Abell}, P.~A., {Allison}, J., {et~al.} 2009,
  arXiv e-prints, arXiv:0912.0201.
\newblock \doarXiv{0912.0201}

\bibitem[{{Masci} {et~al.}(2019){Masci}, {Laher}, {Rusholme}, {Shupe}, {Groom},
  {Surace}, {Jackson}, {Monkewitz}, {Beck}, {Flynn}, {Terek}, {Landry},
  {Hacopians}, {Desai}, {Howell}, {Brooke}, {Imel}, {Wachter}, {Ye}, {Lin},
  {Cenko}, {Cunningham}, {Rebbapragada}, {Bue}, {Miller}, {Mahabal}, {Bellm},
  {Patterson}, {Juri{\'c}}, {Golkhou}, {Ofek}, {Walters}, {Graham}, {Kasliwal},
  {Dekany}, {Kupfer}, {Burdge}, {Cannella}, {Barlow}, {Van Sistine}, {Giomi},
  {Fremling}, {Blagorodnova}, {Levitan}, {Riddle}, {Smith}, {Helou}, {Prince},
  \& {Kulkarni}}]{MasciZTFErrors2019}
{Masci}, F.~J., {Laher}, R.~R., {Rusholme}, B., {et~al.} 2019, \pasp, 131,
  018003, \dodoi{10.1088/1538-3873/aae8ac}

\bibitem[{{Maxted}(2016)}]{ellc}
{Maxted}, P.~F.~L. 2016, {ellc: Light curve model for eclipsing binary stars
  and transiting exoplanets}.
\newblock \doeprint{1603.016}

\bibitem[{McGibbon \& Zhao(2019)}]{CUDAwrapper}
McGibbon, R., \& Zhao, Y. 2019, npcuda-example.
\newblock \url{https://github.com/rmcgibbo/npcuda-example}

\bibitem[{McKinney(2010)}]{mckinney-proc-scipy-2010}
McKinney, W. 2010, in Proceedings of the 9th Python in Science Conference, ed.
  S.~van~der Walt \& J.~Millman, 56--61, \dodoi{10.25080/Majora-92bf1922-00a}

\bibitem[{{McMillan}(2011)}]{McMillan2011GalPop}
{McMillan}, P.~J. 2011, \mnras, 414, 2446,
  \dodoi{10.1111/j.1365-2966.2011.18564.x}

\bibitem[{McNeill {et~al.}(2020)McNeill, Mardling, \&
  M{\"{u}}ller}]{McNeill2020_GW_tides_DWD}
McNeill, L.~O., Mardling, R.~A., \& M{\"{u}}ller, B. 2020, \mnras, 491, 3000,
  \dodoi{10.1093/mnras/stz3215}

\bibitem[{{Mestel}(1952)}]{Mestel1952WDs}
{Mestel}, L. 1952, \mnras, 112, 583, \dodoi{10.1093/mnras/112.6.583}

\bibitem[{{Nelemans} \& {van Haaften}(2013)}]{2013nelemansUCB}
{Nelemans}, G., \& {van Haaften}, L. 2013, in Astronomical Society of the
  Pacific Conference Series, Vol. 470, 370 Years of Astronomy in Utrecht, ed.
  G.~{Pugliese}, A.~{de Koter}, \& M.~{Wijburg}, 153.
\newblock \doarXiv{1302.0878}

\bibitem[{Nickolls {et~al.}(2008)Nickolls, Buck, Garland, \& Skadron}]{CUDA}
Nickolls, J., Buck, I., Garland, M., \& Skadron, K. 2008, Queue, 6, 40,
  \dodoi{10.1145/1365490.1365500}

\bibitem[{{Nissanke} {et~al.}(2012){Nissanke}, {Vallisneri}, {Nelemans}, \&
  {Prince}}]{2012nissanke}
{Nissanke}, S., {Vallisneri}, M., {Nelemans}, G., \& {Prince}, T.~A. 2012,
  \apj, 758, 131, \dodoi{10.1088/0004-637X/758/2/131}

\bibitem[{{Peters} \& {Mathews}(1963)}]{Peters1963}
{Peters}, P.~C., \& {Mathews}, J. 1963, Physical Review, 131, 435,
  \dodoi{10.1103/PhysRev.131.435}

\bibitem[{Tauris \& Konar(2001)}]{Tauris:2001cy}
Tauris, T.~M., \& Konar, S. 2001, Astron. Astrophys., 376, 543,
  \dodoi{10.1051/0004-6361:20010988}

\bibitem[{{The pandas development team}(2020)}]{reback2020pandas}
{The pandas development team}. 2020, {pandas-dev/pandas: Pandas},  Zenodo,
  \dodoi{10.5281/zenodo.3509134}

\bibitem[{Walt {et~al.}(2011)Walt, Colbert, \& Varoquaux}]{Numpy}
Walt, S. v.~d., Colbert, S.~C., \& Varoquaux, G. 2011, Computing in Science and
  Engg., 13, 22, \dodoi{10.1109/MCSE.2011.37}

\bibitem[{Woods {et~al.}(2012)Woods, Ivanova, van~der Sluys, \&
  Chaichenets}]{Woods2012DWD_common_env}
Woods, T.~E., Ivanova, N., van~der Sluys, M.~V., \& Chaichenets, S. 2012, \apj,
  744, 12, \dodoi{10.1088/0004-637X/744/1/12}

\end{thebibliography}

\end{document}